\documentclass[pra,twocolumn, aps,amssymb,footinbib,showpacs,superscriptaddress]{revtex4-1}

\pdfoutput=1

\usepackage[makeroom]{cancel}
\usepackage{color}
\usepackage{graphicx}
\usepackage{amsmath}
\usepackage{amssymb}
\usepackage{xspace}
\usepackage[small]{subfigure}

\newlength{\fighskip} \fighskip=2pt
\newlength{\figvskip} \figvskip=3pt

\newcommand*{\figbox}[2]{{
 \def\figscale{#1}
 \def\arraystretch{0.8}
 \arraycolsep=0pt
 \begin{array}{c}
\vbox{\vskip\figscale\figvskip
  \hbox{\hskip\figscale\fighskip
    \includegraphics[scale=\figscale]{#2}}}
 \end{array}}}

\usepackage{mciteplus} 
\usepackage{dcolumn}
\usepackage{bm}
\usepackage{verbatim}
\usepackage{amscd}
\usepackage{amsfonts}
\usepackage{setspace}
\usepackage{amsthm}
\usepackage{enumerate}
\usepackage{mathtools}

\theoremstyle{plain}

\theoremstyle{plain}

\theoremstyle{plain}
 
\theoremstyle{plain}

\theoremstyle{remark}

\theoremstyle{conjecture}

\theoremstyle{observation}

\theoremstyle{definition}

\theoremstyle{corollary}

\theoremstyle{definition}

\theoremstyle{definition}

\theoremstyle{result}

\theoremstyle{assumption}

\theoremstyle{definition}

\theoremstyle{problem}

\theoremstyle{fact}

\DeclareMathOperator{\OTOC}{OTOC}

\DeclareMathOperator{\Tr}{Tr}

\newcommand{\BY}[1]{{\sl{\textcolor{red}{#1}}}}

\begin{document}

\title{\bf Disentangling Scrambling and Decoherence via Quantum Teleportation}
\author{Beni~Yoshida}
\affiliation{Perimeter Institute for Theoretical Physics, Waterloo, Ontario N2L 2Y5, Canada}
\author{Norman~Y.~Yao}
\affiliation{Department of Physics, University of California Berkeley, Berkeley, California 94720, USA}
\affiliation{Materials Science Division, Lawrence Berkeley National Laboratory, Berkeley, California 94720, USA}

\date{\today}

\begin{abstract}
Out-of-time-order correlation (OTOC) functions provide a powerful theoretical tool for diagnosing chaos and the scrambling of  information in strongly-interacting, quantum systems. However, their direct and unambiguous experimental measurement remains an essential challenge.
At its core, this challenge arises from the fact that the effects of both decoherence and experimental noise can mimic that of information scrambling, leading to decay of OTOCs.
Here, we analyze a quantum teleportation protocol that explicitly enables one to differentiate between  scrambling and decoherence.
Moreover, we demonstrate that within this protocol, one can extract a precise ``noise'' parameter which quantitatively captures the non-scrambling induced decay of OTOCs. 
Using this  parameter, we prove explicit bounds on the true value of the OTOC.
Our results open the door to experimentally measuring quantum scrambling with built-in verifiability. 
\end{abstract}

\maketitle



\section{Introduction}

The thermalization of strongly-interacting systems causes  information about the initial configuration  to become ``scrambled'' at late times, wherein two initial states (with the same conserved quantities) become indistinguishable without measuring a macroscopic number of observables \cite{deutsch1991quantum,srednicki1994chaos,tasaki1998quantum,rigol2008thermalization}. 
Recent studies on the dynamics of such information scrambling have sharpened our understanding of chaos in quantum many-body systems \cite{Hayden07, Sekino08, Lashkari13, Kitaev:2014t1, Maldacena:2016ac, Shenker:2013pqa, Traversable2017, Roberts:2014isa, Shenker:2013yza, Roberts:2014ifa, Hosur:2015ylk, Roberts:2017aa, Blake:2016wvh, Blake:2016aa, Kitaev:2014t2, Nahum:2017aa, Keyserlingk:2017, Cotler:2017aa, Khemani:2017aa, Davison:2017aa, Gu:2017aa, Gao:2016aa}, and have led to new insights on a variety of questions ranging from  the black hole information paradox \cite{Page93,Hayden07, Hosur:2015ylk, Traversable2017, Yoshida:2017aa} to transport phenomena in non-Fermi liquids \cite{banerjee2017solvable,patel2017quantum}.
While a precise definition of quantum scrambling remains elusive, a powerful  proxy for characterizing its behavior is provided by out-of-time order correlation (OTOC) functions, which take the general form: $\langle V(0)W(t)V(0)W(t)\rangle$, where $V, W$ are operators that act on sufficiently small subsystems \cite{larkin1969quasiclassical,Kitaev:2014t1,Shenker:2013pqa,Roberts:2014isa}. 
The intuition behind this correlator is an attempt to measure the influence of one observable at earlier times on another observable at later times --- in essence, a quantum version of the so-called butterfly effect. 
To do this however,  requires the precise reversal of time evolution and thus, poses a daunting challenge for any  experiment.
\begin{figure}
\centering\includegraphics[width=1.\linewidth]{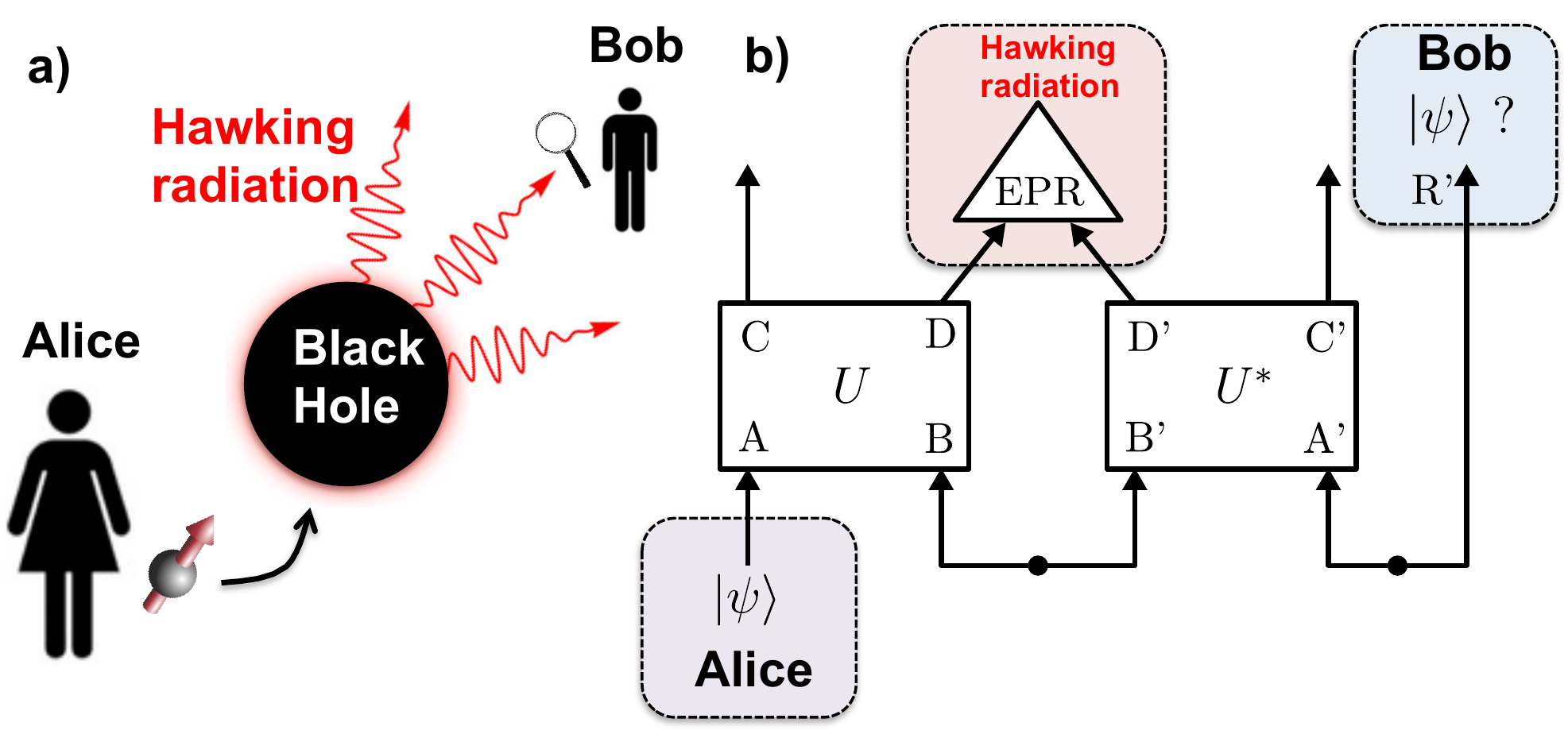}
\caption{a) In the Hayden-Preskill variant of the black-hole decoding problem \cite{Hayden07}, one asks whether Bob can decode the state of Alice's  quantum spin using only Hawking radiation and an entangled partner of the black-hole. Any successful  ``decoding''  serves as affirmation for the existence of  scrambling dynamics. b) For an arbitrary unitary $U$, one can utilize a teleportation-based decoding protocol  to probe the scrambling behavior of the circuit \cite{Yoshida:2017aa}. Crucially, even in the presence of arbitrary noise and imperfections, the teleportation fidelity acts as a metric for quantum scrambling and enables the bounding of the mutual information between Alice and Bob's reference quantum registers. 
}
\label{fig-setup}
\end{figure}

Despite this challenge, a tremendous amount of interest has been devoted to the development of protocols \cite{swingle2016measuring,yao2016interferometric,zhu2016measurement} and platforms \cite{garttner2017measuring,li2017measuring} for the direct measurement of OTOCs. 
The crucial difficulty in interpreting such measurements can be summarized as follows:
For a generic interacting system without symmetries, the scrambling of quantum information will cause out-of-time order correlation functions to decay to zero.
However, both decoherence and imperfect experimental controls (e.g.~time reversal) will \emph{also} cause OTOCs to decay to zero. 
At present, the only way to distinguish between these two contributions --- namely, true chaotic scrambling versus noise and decoherence  --- is to perform full quantum tomography on the many-body system,  requiring exponentially many measurements in the number of qubits \cite{hradil1997quantum,dodonov1997positive,haffner2005scalable,gross2010quantum}.
To this end, the ability to distinguish between genuine quantum information scrambling and extrinsic decoherence remains an essential open question. 


In this paper, we analyze a quantum teleportation protocol that explicitly enables such  differentiation.  
We present three main results. 
First, we demonstrate that within our protocol, one can extract a ``noise'' parameter, which quantifies the \emph{non-scrambling} induced decay of OTOCs. Here, we focus on two illustrative examples: i) depolarization (i.e.~a non-unitary error) and ii) imperfect ``backwards'' time evolution (i.e.~a unitary error).
Second, using this noise parameter, we provide a \emph{bound} on the true scrambling-induced decay of the OTOC.
Again, we analyze two cases, one which applies specifically to the situation of unitary errors and another which applies to arbitrary errors. 
Finally, we describe two simple realizations of our protocol amenable to near-term, intermediate scale qubit and qutrit systems as well as their generalizations to include Grover search \cite{grover1996fast}. 

The essence of our approach is based upon a recent decoding algorithm for the Hayden-Preskill variant of the black hole information problem \cite{Hayden07, Yoshida:2017aa}.
The connection between this decoding algorithm and information scrambling can be understood as follows: If the dynamics of a black hole are unitary, then one should in principle, be able to retrieve a quantum state that is thrown in from the Hawking radiation that comes out (Fig.~1a). Crucially, it turns out such a successful ``decoding'' of the original quantum state serves as smoking-gun evidence for the existence of true scrambling dynamics.

Our manuscript is organized as follows. In Sec.~II, we begin by reviewing the information theoretic interpretation of scrambling and OTOCs. Then, using the example of a depolarizing quantum channel, we illustrate the fact that decoherence can result in the decay of OTOCs even in the absence of scrambling dynamics. This allows us to propose a sharp measure which quantifies the ratio of scrambling-induced versus decoherence-induced OTOC decay. Moreover, it reveals that the genuine metric for scrambling should be taken as the mutual information between subsystems and not simply the measured OTOC.
The groundwork being laid, in Sec.~III, we introduce the teleportation-based decoding protocol and clarify its operation in the ideal case without noise and decoherence. Then in Sec.~IV, we turn to an analysis of the protocol in the presence of \emph{arbitrary} noise and decoherence. Here, we demonstrate that the protocol provides a quantitative estimate for the amount of dissipation in the system.  In Sec.~V, motivated by recent experiments, we restrict ourselves to a sub-class of noise and imperfections, with a focus on \emph{coherent} errors. Under this restriction, we show that one can explicitly bound the ideal value of the OTOC (i.e. in the absence of errors), using the experimentally \emph{measured} value of the  OTOC. 
In Sec.~VI, we generalize such a bound to the case of arbitrary errors and prove that one can utilize the teleportation fidelity to  bound the mutual information between subsystems (and hence the amount of scrambling). 
Finally, in Sec.~VII, we propose and analyze two experimental implementations of our protocol in near-term intermediate scale quantum simulators. We focus on a class of Clifford scramblers that  
saturate the lower bound for OTOCs.
In Sec.~VIII, we offer some concluding remarks and intriguing directions to be pursued.

%

%

%
%
%
%
%
%


\section{Characterizing scrambling and decoherence}

\subsection{Definition of scrambling in terms of OTOCs}

Let us begin by providing a definition for quantum scrambling in terms of the behavior of  out-of-time order correlation  functions~\cite{Roberts:2017aa, Yoshida:2017aa}: 
\begin{align}
\langle O_{X}O_{Y}(t)O_{Z}O_{W}(t) \rangle \approx \langle O_{X}O_{Z}\rangle\langle O_{Y} \rangle\langle O_{W} \rangle + \nonumber \\ \langle O_{X}\rangle \langle O_{Z}\rangle\langle O_{Y}  O_{W} \rangle - \langle O_{X}\rangle \langle O_{Y}\rangle\langle O_{Z} \rangle\langle O_{W} \rangle \label{eq:def}
\end{align}
where $O_{X},O_{Z}$ are operators that act on sub-system $A$ (at time zero) and  $O_{Y},O_{W}$ are operators that act on sub-system $D$ (at time $t$), as depicted in Fig.~\ref{fig-setup}.
%
This equation becomes exact in the thermodynamic limit for chaotic systems at late times and can also be derived from the eigenstate thermalization hypothesis  \cite{deutsch1991quantum,srednicki1994chaos,tasaki1998quantum,rigol2008thermalization,Huang:2017aa}. 
While we will focus on infinite temperature systems with $\rho=\frac{1}{d}\mathbb{I}$, we note that this definition naturally generalizes to finite temperatures. 
Our above definition of scrambling is required to hold for \emph{all} local operators, but a slightly a more coarse-grained characterization of scrambling (and one which is easier to probe experimentally) can be achieved via the averaged OTOC~\cite{Hosur:2015ylk}: 
\begin{align}
\overline{\langle \OTOC \rangle} \equiv \iint d O_{A} d O_{D} \langle O_{A} O_{D}(t)O_{A}^{\dagger} O_{D}^{\dagger}(t)\rangle  
\end{align}
where $\int d O_{R}$ is the Haar-average over all unitary operators  on sub-system $R$.  This Haar integral can  be replaced by an average over Pauli operators:
\begin{align}
\iint d O_{A} d O_{D} \langle O_{A} O_{D}(t)O_{A}^{\dagger} O_{D}^{\dagger}(t)\rangle = \nonumber \\ \frac{1}{d_{A}^2d_{D}^2}\sum_{P_{A},P_{D}} \langle P_{A} P_{D}(t)P_{A}^{\dagger} P_{D}^{\dagger}(t)\rangle \label{eq:1-design},
\end{align}
where $P_{A(D)}$ are Pauli operators and $d_{A(D)}$ is the dimension of the sub-system \footnote{Eq.~\eqref{eq:1-design} holds since the Pauli operators form a unitary $1$-design \cite{renes2004symmetric,dankert2009exact}. Note that there are $d_{A}^2,d_{D}^2$ Pauli operators (including the identity operator) on regions $A,D$, respectively.}.
Working at infinite temperature and using Eqn.~\eqref{eq:def} then yields the  scrambled value of the averaged OTOC as \footnote{We note that there are unitary operators which satisfy Eq.~\eqref{eq:ave-def}, but not Eq.~\eqref{eq:def}. For example, a random Clifford operator is scrambling for Eq.~\eqref{eq:ave-def}, since the Clifford operators form a unitary $2$-design. However, OTOCs for a Clifford unitary are always $\pm1$ if $O_{X}=O_{Z}$ and $O_{Y}=O_{W}$ are Pauli operators, and thus do not satisfy Eq.~\eqref{eq:def}. In this sense, a random unitary from a $2$-design is not enough to achieve full scrambling. Rather, to achieve full scrambling, it suffices to pick a random operator $U$ from a unitary $4$-design.}:
\begin{align}
\overline{\langle \OTOC \rangle}_{\text{S}}  \approx \frac{1}{d_{A}^2} + \frac{1}{d_{D}^2} - \frac{1}{d_{A}^2d_{D}^2} \label{eq:ave-def}.
\end{align}
This scrambled value, $\overline{\langle \OTOC \rangle}_{\text{S}}$, is achieved for a Haar random unitary as $d\rightarrow \infty$ \cite{Hosur:2015ylk}.
On the other hand, for  arbitrary unitary time evolution, $\overline{\langle \OTOC \rangle}$ is bounded from above by unity and from below by $\max(\frac{1}{d_{A}^2}, \frac{1}{d_{D}^2}  )$; the fact that it never fully decays to zero is because it contains contributions from cases where  $P_{A}=\mathbb{I}$ or $P_{D}=\mathbb{I}$.
We note that the minimal scrambled value is only asymptotically achieved  for large systems with $d \gg d_{A}\gg d_{D}$ or $d \gg d_{D}\gg d_{A}$ \cite{suppinfo}.

\subsection{Decoding as a route to scrambling}

In order to characterize the effect of decoherence on the averaged OTOC, it will be useful to first recall the information theoretic interpretation of $\overline{\langle \OTOC \rangle}$ in terms of the mutual information between sub-systems. 
To do so, we will utilize the so-called state representation of the time-evolution operator, $U$ \cite{Hayden07, Hosur:2015ylk}. This  representation allows us to view a unitary operator $U$, acting on an $n$-qubit Hilbert space $\mathcal{H}_{AB}$, as a pure quantum state, supported on a $2n$-qubit Hilbert space $\mathcal{H}_{AB}\otimes \mathcal{H}_{RB'} (\simeq \mathcal{H}_{RCDB'} )$:
\begin{align}
\hspace{-10mm} |\Psi\rangle \equiv (I_{R}\otimes U_{AB}\otimes I_{B'}) |\text{EPR}\rangle_{RA}\otimes |\text{EPR}\rangle_{BB'} = \nonumber \hspace{-10mm}  \\ 
\figbox{1.0}{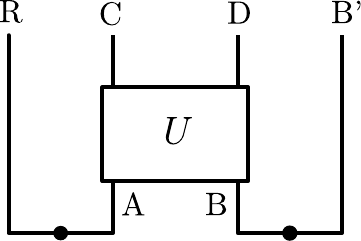}\qquad  \label{eq:world-state}
\end{align}
where time runs upward and the horizontal lines in the diagram represent EPR pairs ($|\text{EPR}\rangle \equiv\frac{1}{\sqrt{d}}\sum_{j=1}^{d} |j\rangle \otimes |j\rangle$), while the dots capture the $\frac{1}{\sqrt{d}}$ normalization factor   in the EPR pair.  
%
Crucially, this representation allows us to characterize the scrambling behavior of the time evolution, $U$, via  the entanglement properties of the pure state, $|\Psi\rangle$!

Three remarks are in order.
First, for non-interacting time evolutions, including  free-fermion dynamics or SWAP operators, $|\Psi\rangle$ contains mostly bipartite entanglement among subsystems. 
On the other hand, for strongly-interacting time evolutions that lead to scrambling, $|\Psi\rangle$ consists of \emph{multipartite} entanglement delocalized over the full Hilbert space  $RCDB'$.

Second, we note that $|\Psi\rangle$ is precisely the state of interest in the Hayden-Preskill thought experiment \cite{Hayden07}. 
In particular, the Hilbert spaces $A,B,C,D$  [Eqn.~\eqref{eq:world-state}] support, respectively, Alice's input states, the initial black hole, the remaining black hole and the Hawking radiation. Meanwhile, $R$ serves as a reference  for Alice's input state, while $B'$ is the entangled partner of the black hole.
The Hayden-Preskill decoding problem can then be stated as follows: when can Bob decode Alice's quantum state using only the Hawking radiation $D$ and the entangled black-hole partner $B'$. 
The answer, somewhat naturally, is when the (von Neumann) mutual information between $R$ and $B'D$ is maximal.
More precisely, when this is the case, there  exists a unitary operator acting on $B'D$ which distills an EPR pair between $R$ and $B'D$ with high fidelity, thereby faithful recovering Alice's input state~\cite{Hayden07, Hayden:2008aa}.

\begin{figure}
\centering\includegraphics[width=0.55\linewidth]{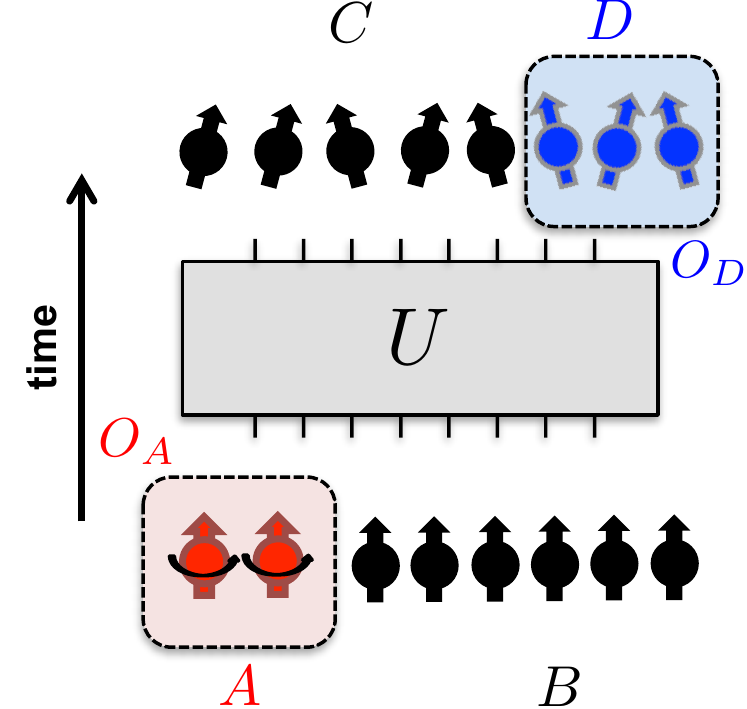}
\caption{Schematic image of the setup associated with an out-of-time ordered correlation function, $\langle O_{A}(0)O_{D}(t)O_{A}(0)O_{D}(t) \rangle$. Time runs upward and evolution is generated by the unitary $U$.  The Hilbert space decomposes as $\mathcal{H}=\mathcal{H}_{A}\otimes \mathcal{H}_{B}=\mathcal{H}_{C}\otimes \mathcal{H}_{D}$. Ideally, operators $O_{A}$ and $O_{D}$ act on sufficiently small subsystems of the full system. 
}
\label{fig-setup}
\end{figure}

Third, we note that for maximally mixed states the R\'{e}nyi-$2$ mutual information, $I^{(2)}(R,B'D)$, lower bounds the von Neumann mutual information, $I(R,B'D)$ \footnote{It is possible to generalize our results to finite temperature (factorable ensembles) using the R\'{e}nyi divergence \cite{suppinfo}.}.~This is particularly useful since  $I^{(2)}(R,B'D)$ is in fact, directly related to our previously defined averaged OTOC~\cite{Hosur:2015ylk, Roberts:2017aa}: 
\begin{align}
\overline{\langle \OTOC \rangle} = 2^{ - I^{(2)}(R,B'D) }, \label{eq:OTOC-unitary}
\end{align}
where $I^{(2)}(R,B'D) \equiv S_{R}^{(2)} + S_{B'D}^{(2)} - S_{RB'D}^{(2)}$  and the R\'{e}nyi-$2$ entropies, $S^{(2)}$, are evaluated with respect to the  state $|\Psi\rangle$. To this end, in an \emph{ideal} (noise-less) system, the smallness of $\overline{\langle \OTOC \rangle}$, which characterizes the amount of scrambling in the system, is \emph{also} sufficient to diagnose Bob's faithful recovery of Alice's state
\footnote{While the R\'{e}nyi-$2$ mutual information is a measurable quantity as the average of OTOCs, the standard mutual information ($\alpha=1$) is often more convenient as it satisfies useful monotonicity inequalities. For the case of maximally mixed ensembles $\rho=\frac{1}{d}I$, one can derive $I(A,B'D)\geq I^{(2)}(A,B'D)$ using the monotonicity of R\'{e}nyi entropy. This analysis can be generalized to cases where the input and output ensembles factorize \cite{suppinfo}; $\rho_{AB}=\rho_{A}\otimes \rho_{B}$ and $\rho_{CD}=\rho_{C}\otimes \rho_{D}$ where the R\'{e}nyi-$2$ mutual information is replaced with a certain expression involving the R\'{e}nyi-$2$ divergence from which the standard mutual information can be lower bounded. See appendix for details. 
}.

However, the essential point is that in a system with noise and imperfections, the smallness of $\overline{\langle \OTOC \rangle}$ can either result from decoherence or from true scrambling behavior. Crucially, only the latter will contribute to Bob's ability to decode Alice's quantum state! 
In the following subsections, we will first focus on identifying the effects of decoherence on the averaged OTOC. With this in hand, we will then provide a precise metric to distinguish between decoherence and scrambling.

\subsection{Effects of decoherence on the OTOC}

To understand the effects of decoherence, let us consider the following quantum channel $\mathcal{Q}$:
\begin{align}
\rho \rightarrow \mathcal{Q}(\rho) = (1-p) U \rho U^{\dagger} + p \frac{1}{d} \Tr(\rho) \label{eq:noise}
\end{align}
which suffers from  depolarization with probability $p$ \cite{nielsen2010quantum}. 
For traceless operators, one finds that the out-of-time order correlators behave as:
\begin{align}
\langle O_{X}\widetilde{O_{Y}}(t)O_{Z}\widetilde{O_{W}}(t) \rangle = (1-p)^2 \langle O_{X}O_{Y}(t)O_{Z}O_{W}(t) \rangle
\label{eq:OTOCdefn}
\end{align}
where we use the tilde to indicate observables time-evolved under the quantum channel $\mathcal{Q}$, while time-evolved operators without a tilde are evolved under the unitary portion of the channel, $U$ \footnote{We note that in the presence of arbitrary forms of noise and decoherence, the experimentally ``measured'' value of the OTOC may depend on the specific measurement protocol.  For example, the OTOC measured via interferometric protocols \cite{swingle2016measuring,yao2016interferometric,zhu2016measurement,garttner2017measuring,li2017measuring} will generically differ from the OTOC measured via our teleportation protocol. However, for the important case of a purely depolarizing channel as per Eqn.~\eqref{eq:noise}, all such protocols will measure the same OTOC given by Eqn.~\eqref{eq:OTOCdefn}. Moreover, while the quantitative values of noisy OTOCs may differ between protocols, their qualitative decay in the presence of decoherence is generic. To this end, a key difference between our decoding protocol and previously proposed interferometric protocols is the initial preparation of EPR pairs; this preparation is not present in the case of interferometric protocols and underlies the reason why our teleportation-based method can verify the existence of scrambling dynamics while prior methods cannot.  }.
Thus, even in the absence of information scrambling (i.e.~in the actual behavior of $\langle O_{X}O_{Y}(t)O_{Z}O_{W}(t) \rangle$), the \emph{measured} OTOCs for the channel $\mathcal{Q}$, can  become small owing to decoherence; in particular, undergoing depolarization with a finite probability per unit time induces an exponential decay of the measured values of OTOCs. 

The difference between scrambling and decoherence can be further sharpened and made precise by considering the late-time asymptotics of OTOCs, which serve as our operational definition of quantum scrambling in Eqn.~\eqref{eq:def}. Specifically,  under a completely depolarizing channel (e.g.~$p=1$), the out-of-time order correlators decompose as follows:
\begin{align}
\langle O_{X}O_{Y}(t)O_{Z}O_{W}(t) \rangle = \langle O_{X}O_{Z}\rangle\langle O_{Y} \rangle\langle O_{W} \rangle,
\end{align}
which contains only the first term in Eqn.~\eqref{eq:def}.

As before, one can also examine the averaged OTOCs associated with the channel $\mathcal{Q}$:
\begin{align}
\langle \widetilde{\OTOC} \rangle \equiv \iint d O_{A} dO_{D} \langle O_{A}\widetilde{O_{D}}(t)O_{A}^{\dagger}\widetilde{O_{D}^{\dagger}}(t) \rangle. 
\label{eq:OTOCtildechannel}
\end{align}
Note that for an arbitrary quantum channel, the value of $\langle \widetilde{\OTOC} \rangle$ is now lower bounded by $\min(\frac{1}{d_{A}^2},\frac{1}{d_{D}^2})$, whereas in the absence of imperfections, $\langle \overline{\OTOC} \rangle$ was previously lower bounded by $\max(\frac{1}{d_{A}^2},\frac{1}{d_{D}^2})$.

In the above discussion, we have implicitly assumed that both $O_{Y}$ and $O_{W}$ are evolved with the same (possibly imperfect) quantum channel $\mathcal{Q}$. 
However, it is certainly of interest to consider the situation where they evolve under two different quantum channels, which is precisely the experimental scenario if one performs backwards time evolution imperfectly. 
We will address this case in detail a bit later. 

\subsection{Distinguishing decoherence from scrambling}

As we have shown, for an arbitrary quantum channel, the decay of OTOCs is not sufficient to experimentally diagnose the scrambling behavior of the system.
To this end, we now provide a formal metric for distinguishing between scrambling and decoherence in noisy quantum systems.
Let us  consider the state representation of the channel $\mathcal{Q}$ defined as follows:
\begin{align}
\hspace{-10mm} \rho \equiv \mathcal{Q}(|\text{EPR}\rangle \langle\text{EPR}|_{RA}  \otimes |\text{EPR}\rangle \langle \text{EPR}|_{BB'}) = \nonumber \hspace{-10mm} \\ \figbox{0.9}{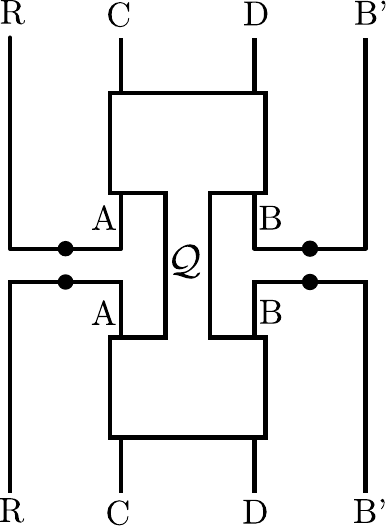}\ ,
\end{align}
where $\rho$ is the system's density matrix. 

To gain some intuition, let us consider the two limiting cases: $p=0$ (no decoherence) and $p=1$ (full depolarization). In the first case, $\mathcal{Q}$ is purely unitary and can be decomposed into two separate boxes corresponding to $U$, $U^{\dagger}$, wherein $\rho = |\Psi \rangle \langle \Psi|$ is a pure state with $|\Psi \rangle$ as defined in Eqn.~\eqref{eq:world-state}. 
In the second case,  $\mathcal{Q}$ induces  complete depolarization and the corresponding quantum state, $\rho$, is a maximally mixed state on $RCDB'$ with graphical representation:
\begin{align}
\rho = \frac{1}{d^2} \mathbb{I}_{R}\otimes  \mathbb{I}_{C}\otimes \mathbb{I}_{D}\otimes \mathbb{I}_{B'}= \ \figbox{0.8}{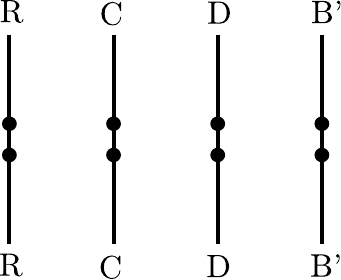}\ .
\end{align}
%
As one can see from this example, for non-unitary time evolution, $\rho$ is \emph{not} a pure state. 

However,  one nevertheless finds that the averaged OTOC can be re-expressed in terms of R\'{e}nyi-$2$ entropies  evaluated with respect to $\rho$ \footnote{
We do not know the terminology for the expression $S_{BD}^{(2)} + S_{D}^{(2)} - S_{B}^{(2)}$, but it is worth noting that $S_{BD} + S_{D} - S_{B} \geq 0$ corresponds to the celebrated Araki-Lieb inequality. 
}:
\begin{align}
\langle \widetilde{\OTOC} \rangle= 2^{- (S_{B'D}^{(2)} + S_{D}^{(2)} - S_{B'}^{(2)}) }.
\label{eq:OTOCtilde_entropy}
\end{align}
The astute reader may wonder why this looks quite similar to the aforementioned result in the ideal, noiseless case [Eqn.~\eqref{eq:OTOC-unitary}]?
Since $S_{R}^{(2)}+S_{B'}^{(2)}=S_{C}^{(2)}+S_{D}^{(2)}=n$ (where $n$ is the total number of  qubits in $RB'$), if $\rho$ was in fact a pure state, then one would have $S_{RB'D}^{(2)} = S_{C}^{(2)}$ and hence:
\begin{align}
S_{B'D}^{(2)} + S_{D}^{(2)} - S_{B'}^{(2)}= I^{(2)}(R,B'D).
\end{align}
Thus, when $\mathcal{Q}$ is unitary (e.g.~when the depolarizing probability $p=0$), the averaged OTOC indeed reduces to our previous result for the ideal system [Eqn.~\eqref{eq:OTOC-unitary}]. 

Crucially, for a generic noisy quantum channel, the state $\rho$ is \emph{not} pure and $S_{B'D}^{(2)} + S_{D}^{(2)} - S_{B'}^{(2)} \neq I^{(2)}(R,B'D)$!
Herein lies the essence of our result: The genuine metric for scrambling, the mutual information, is \emph{not} directly measured via the OTOC, which instead only measures the entropy, $S_{B'D}^{(2)} + S_{D}^{(2)} - S_{B'}^{(2)}$.

The deviation between these two quantities serves as a natural metric or ``noise parameter'' capturing the decoherence present in   the channel $\mathcal{Q}$:
\begin{align}
\delta \equiv \frac{2^{I^{(2)}(R,B'D)}}{2^{S_{B'D}^{(2)} + S_{D}^{(2)} - S_{B'}^{(2)}}} = 2^{S_{C}^{(2)} -S_{RB'D}^{(2)} },
\label{eq:deltanoiseparam}
\end{align}
where $\delta=1$ for unitary time evolution while $\delta = 1/d_{D}^2$ for a completely depolarizing channel.
Note that for  any $\delta < 1$, one knows that decoherence is at least partially responsible for the observed decay in the averaged OTOC. 
More succinctly, there are two physical mechanisms that  cause $\langle \widetilde{\OTOC}\rangle$ to decay.  First, entangling  $B'D$ with $R$ (as per unitary scrambling) and second,  entangling $B'D$ with the environment (as in a depolarizing channel); $\delta$ captures the ratio between these two contributions.  

In the following sections, we will turn to the experimental measurement and characterization of $\delta$, via a quantum teleportation decoding-protocol \cite{Hayden07, Yoshida:2017aa}. In Sec.~III, we will begin by setting up the framework of the  protocol in the ideal case (decoherence and noise free), while in Sec.~IV, we will shift our attention to investigate a variety of  imperfections (i.e.~both unitary and non-unitary errors).

\section{Teleportation-based Decoding Protocol (ideal case)}

\subsection{Representing the OTOC as a thermofield double state}

%
To begin, let us consider the diagrammatic representation of the OTOC in the case of unitary time-evolution $U$:
\begin{align}
\langle O_{A}O_{D}(t)O_{A}^{\dagger}O_{D}^{\dagger}(t) \rangle =\ \figbox{0.7}{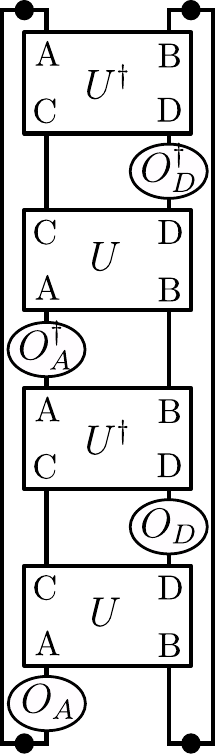} \label{eq:OTOC-unfold}
\end{align}
where again, time runs upward and the expectation value is taken at infinite temperature; in our diagrammatic representation, connecting the legs of the input and output corresponds to taking a trace with respect to a maximally mixed state. 
While the OTOC, $\langle O_{A}O_{D}(t)O_{A}^{\dagger}O_{D}^{\dagger}(t) \rangle$, is defined on the Hilbert space $\mathcal{H}_{AB}$, it can be recast as the  expectation value of  local operators on the doubled Hilbert space $\mathcal{H}_{AB}\otimes \mathcal{H}_{B'A'}$. In particular, consider the following state, $|\Phi_{O_{A}}\rangle \equiv (U_{AB}\otimes U_{B'A'}^{*})  (O_{A}\otimes I_{BB'A'}) | \text{EPR} \rangle_{ABB'A'}$, which lives in $\mathcal{H}_{AB}\otimes \mathcal{H}_{B'A'}$.
This is the  so-called thermofield double state (at infinite temperature)  perturbed by local operator  $O_{A}$ and then time-evolved by $U\otimes U^{*}$. Taking the expectation value of $I_{C}\otimes O_{D} \otimes O_{D}^{*}\otimes I_{C'}$ in this state results in:
\begin{align}
\hspace{-10mm}  \langle \Phi_{O_{A}} | I_{C}\otimes O_{D} \otimes O_{D}^{*}\otimes I_{C'}| \Phi_{O_{A}} \rangle = \nonumber  \hspace{-10mm}  \\ \figbox{0.9}{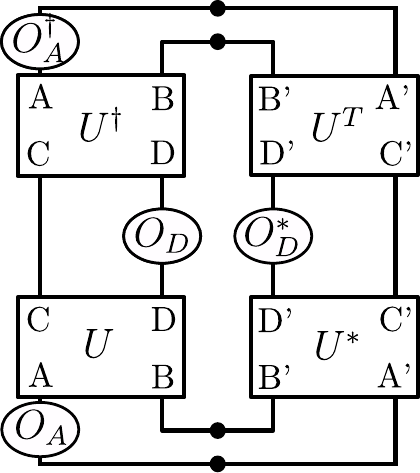} \label{eq:OTOC}
\end{align}
which is exactly equivalent to the OTOC defined in Eqn.~\eqref{eq:OTOC-unfold}. This equivalence is most easily seen by ``unfolding'' the diagram of Eqn.~\eqref{eq:OTOC-unfold} while noting that $(U\otimes I)|\text{EPR}\rangle=(I\otimes U^{T})|\text{EPR}\rangle$, or in diagrammatic form:
\begin{align}
\figbox{1.0}{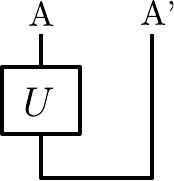} = \figbox{1.0}{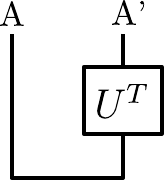}\ .
\end{align}
Since  $\langle O_{A}O_{D}(t)O_{A}^{\dagger}O_{D}^{\dagger}(t) \rangle = \langle \Phi_{O_{A}} | I_{C}\otimes O_{D} \otimes O_{D}^{*}\otimes I_{C'}| \Phi_{O_{A}} \rangle$, one can directly measure OTOCs as an expectation value of $O_{D} \otimes O_{D}^{*}$ in the doubled Hilbert space $\mathcal{H}_{AB}\otimes \mathcal{H}_{B'A'}$; then to compute $\overline{\langle\OTOC \rangle}$, one can  simply average over the various operators: $O_{A}, O_{D}$. 

As aforementioned, a more elegant and efficient method for measuring $\overline{\langle\OTOC \rangle}$ has recently emerged in the form of a probabilistic decoding protocol (via postselected teleportation) for the Hayden-Preskill thought experiment \cite{Yoshida:2017aa}. 

\subsection{Decoding protocol in the ideal case}

In the decoding protocol, in addition to Alice's reference state, Bob also prepares an additional EPR pair $|\text{EPR}\rangle_{A'R'}$ before applying $U^{*}$ to both the entangled black-hole partner $B'$ and the $A'$-part of his EPR pair. 
In order to decode Alice's state, Bob must create an EPR pair between Alice's reference state $R$ and his remaining register qubit, $R'$. 
After  time evolution, the system is in the state:
\begin{equation}
\begin{split}
|\Psi_{\text{in}}\rangle &=(I_{R}\otimes U_{AB} \otimes U^{*}_{B'A'}\otimes I_{R'}) \\ &\hspace{5mm}|\text{EPR}\rangle_{RA}\otimes|\text{EPR}\rangle_{BB'}\otimes|\text{EPR}\rangle_{A'R'} \\
&= \figbox{1.0}{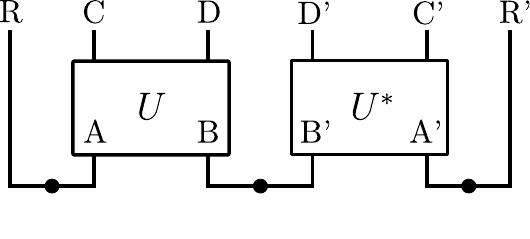}\ .
\end{split}
\end{equation}
Next, Bob collects pairs of qubits on $DD'$ and performs a projective measurement onto $|\text{EPR}\rangle_{DD'}$, resulting in the state:
\begin{align}
|\Psi_{\text{out}}\rangle = \frac{1}{\sqrt{P_{\text{EPR}}}} I_{RC}\otimes \Pi_{DD'}\otimes I_{C'R'}  |\Psi_{\text{in}}\rangle =\nonumber \\  \frac{1}{\sqrt{P_{\text{EPR}}}}\ \figbox{1.0}{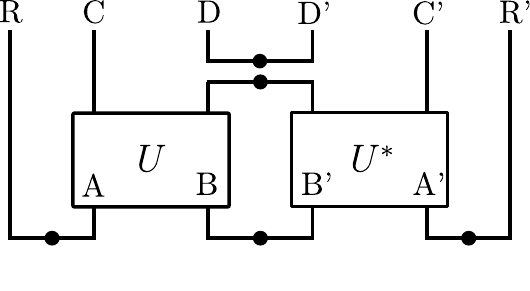}  \label{psiout}
\end{align}
where $P_{\text{EPR}}$ represents the probability of measuring $|\text{EPR}\rangle_{DD'}$. Noting that $\langle \Psi_{\text{out}}|\Psi_{\text{out}}\rangle = \frac{1}{P_{\text{EPR}}} \langle \Psi_{\text{in}}| I_{RC}\otimes \Pi_{DD'}\otimes I_{C'R'} | \Psi_{\text{in}}\rangle = 1$, yields the diagram for $P_{\text{EPR}}$:
\begin{equation}
\begin{split}
P_{\text{EPR}} \equiv \langle \Psi_{\text{in}}| I_{RC}\otimes \Pi_{DD'}\otimes I_{C'R'} | \Psi_{\text{in}}\rangle 
=  \\ \figbox{0.9}{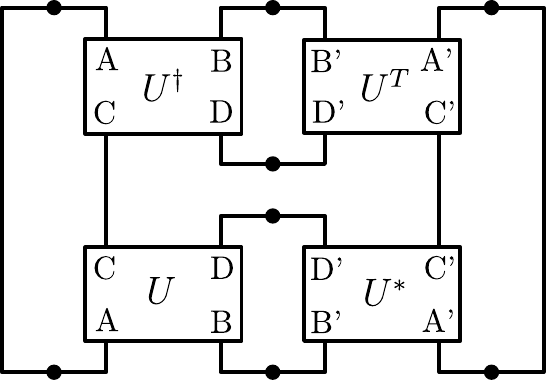} \label{eq-P}\ ,
\end{split}
\end{equation}
where $\Pi_{DD'}=|\text{EPR}\rangle\langle\text{EPR}|_{DD'}$ represents the projective measurement. 
The fidelity of Bob's decoding (of Alice's state) can then be computed via the EPR projection fidelity on $RR'$:
\begin{align}
F_{\text{EPR}} \equiv \langle\Psi_{\text{out}}|I_{CDD'C'} \otimes \Pi_{RR'} |\Psi_{\text{out}}\rangle. 
\end{align}
It has been shown \cite{Yoshida:2017aa} that  if the time-evolution $U$ is scrambling,  an EPR pair $|\text{EPR}\rangle_{RR'}$ can be distilled with high fidelity by post-selecting the measurement result on $|\text{EPR}\rangle_{DD'}$. 
Thus, the projection, $\Pi_{DD'}$, not only serves to decouple Bob's register $R'$ from the remaining black holes, $CC'$, but also teleports Alice's quantum state $|\psi\rangle$ to Bob's register (Fig.~1b). 

In the ideal, noiseless case, this probabilistic decoding protocol enables one to measure the averaged OTOC associated with $U$ in two \emph{different} ways,  using the values of $P_{\text{EPR}}$ and $F_{\text{EPR}}$, respectively.
First, noting that $\int dO_{D}\  O_{D}\otimes O_{D}^{*} = \Pi_{DD'}$, one finds via a simple graphical derivation that~\cite{Roberts:2017aa}, 
\begin{align}
P_{\text{EPR}} = \overline{\langle \OTOC \rangle}.
\end{align}
Thus, by keeping track of the probability associated with the projective measurement, $\Pi_{DD'}$, one directly measures the averaged OTOC.
In the case of $F_{\text{EPR}}$, one can use the following equation:
\begin{equation}
\begin{split}
P_{\text{EPR}}F_{\text{EPR}} &=\langle \Psi_{\text{in}}| \Pi_{RR'}\Pi_{DD'}\otimes I_{CC'} | \Psi_{\text{in}}\rangle \\
&= \frac{1}{d_{A}^2}\ \figbox{0.9}{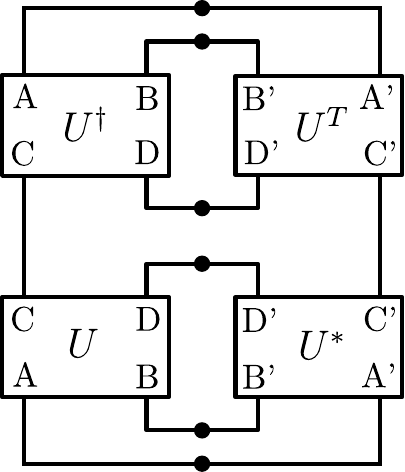} \\ &= \frac{ 1}{d_{A}^2}
\end{split} \label{pf-nonoise}
\end{equation}
to derive
\begin{align}
F_{\text{EPR}} = \frac{1}{d_{A}^2  \overline{\langle \OTOC \rangle}}.
\end{align}
To this end, the teleportation fidelity of Alice's state into Bob's register also directly encodes the averaged OTOC!

While both $P_{\text{EPR}}$ and $F_{\text{EPR}}$ measure $\langle \overline{\OTOC} \rangle$, there is an important (but subtle) distinction from the perspective of experiments; in particular, for a scrambling unitary, the former becomes small while the latter becomes large.
Thus, when using $P_{\text{EPR}}$, an experiment cannot distinguish between a decay in signal arising from scrambling or decoherence. 
On the other hand, when using $F_{\text{EPR}}$, since decoherence can never enhance the fidelity, a successful decoding always serves as a definite signature of quantum scrambling.
This difference will become more apparent in Sec.~IV when we explicitly consider the effects of noise and decoherence.

%

\subsection{Teleportation of a quantum state}

In the previous subsection, we have formulated the decoding protocol in terms of the distillation of EPR pairs on $RR'$. 
This formulation implicitly assumes an average over Alice's input state $|\psi\rangle$. 
However, in the context of experiments, one necessarily perform the teleportation protocol for individual quantum states. 
Moreover, for dynamics that are not fully scrambling, the dependence of the decoding fidelity on the initial state can be used to discern certain properties of the unitary.
An example of this is provided by a system evolving under classical random dynamics, where teleportation only occurs for computational basis states.

To this end, we now consider the decoding protocol for a specific input wavefunction, where Alice prepares $|\psi\rangle$ on $A$, and Bob checks to see if he obtains $|\psi\rangle$ on $R'$ (Fig.~1b):
\begin{align}
\figbox{1.0}{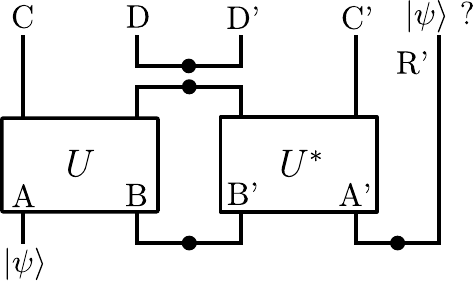}
\end{align}
Interestingly, such a setup for decoding specific states can probe more fine-grained properties of OTOCs. The probability of measuring an EPR pair on $DD'$ is given by
\begin{align}
P_{\psi} =\ \figbox{0.9}{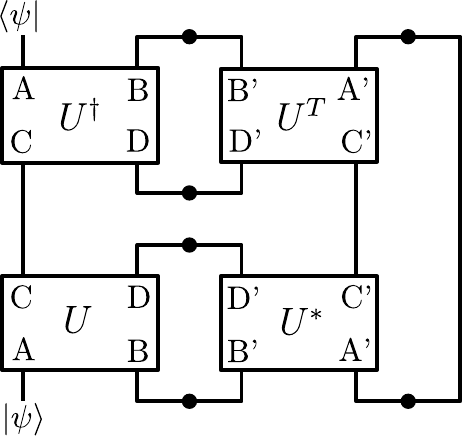}\ .
\end{align}
This probability can be re-expressed in terms of OTOCs as follows:
\begin{align}
P_{\psi} =  \iint d O_{D} d\phi \ \langle O_{A} O_{D}(t)O_{A}^{\dagger}O_{D}^{\dagger}(t)\rangle, 
\end{align}
where $O_{A}= |\psi\rangle \langle \phi |$ and the average over $O_{A}$  is performed by integrating over $|\phi\rangle$. 
It suffices to take an average over any set of orthogonal states (i.e.~$\{|0\rangle,|1\rangle,|2\rangle,\ldots\}$), since the above Haar-integral involves only the first moment of $|\phi\rangle$. By inserting an EPR projection onto  $CC'$, one arrives at the following lower bound,
\begin{align}
P_{\psi}\geq \ \figbox{0.9}{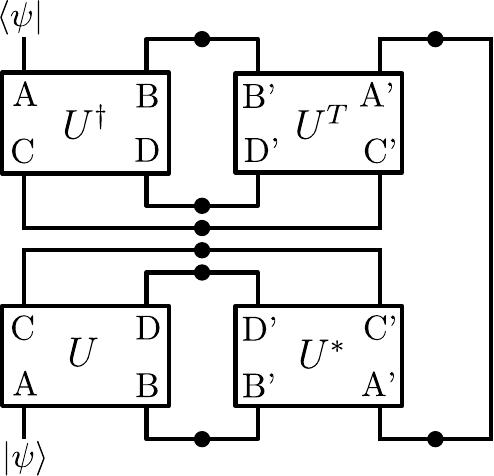}\ = \frac{1}{d_{A}^2}.
\end{align}
To recover $P_{\text{EPR}}$, one simply averages over (orthogonal) states, $P_{\text{EPR}} = \int d\psi \ P_{\psi}$.
Since the minimal value of $P_{\text{EPR}}$ is also $\frac{1}{d_{A}^2}$, this minimum is achieved when $P_{\psi} = \frac{1}{d_{A}^2}$ for all states.
%
Letting $F_{\psi}$ be the decoding fidelity after  postselection, one finds 
\begin{align}
P_{\psi}F_{\psi} = \frac{1}{d_{A}}\ \figbox{1.0}{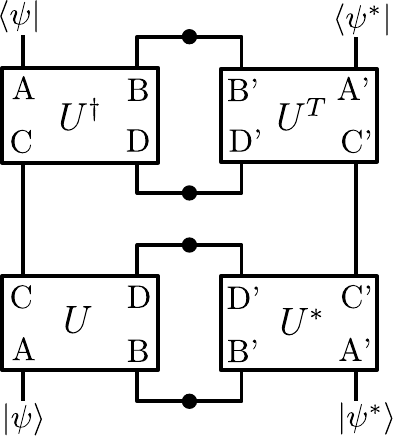}\ .
\end{align}
leading to the  bound \footnote{To derive this lower bound, we again insert an EPR projector on $CC'$ into the diagram for $P_{\psi}F_{\psi}$.},
\begin{align}
P_{\psi}F_{\psi}\geq \frac{1}{d_{A}^2}\quad \Rightarrow \quad F_{\psi} \geq \frac{1}{d_{A}^2P_{\psi}}.
\label{eq:bound-state}
\end{align}
Thus, in the ideal case, a small value of $P_{\psi}$  guarantees the faithful post-selected teleportation of $|\psi\rangle$ from Alice to Bob.
In contrast to the previous subsection, we note that the value of $P_{\psi}F_{\psi}$ depends on the initial state $|\psi\rangle$.

 One can also recast $P_{\psi}F_{\psi}$ as an OTOC,
\begin{align}
P_{\psi}F_{\psi} = \int d O_{D} \langle O_{A} O_{D}(t)O_{A}^{\dagger}O_{D}^{\dagger}(t)\rangle,
\label{eq:pfotocstates}
\end{align}
where $O_{A}= |\psi\rangle \langle \psi |$.  Then, by averaging  over  input states \footnote{This integral over $|\psi \protect \rangle$  can be replaced with an average over a set of states that form a $2$-design. One example is the set of eigenstates of the Pauli operators.}, one obtains
\begin{align}
\int d\psi P_{\psi}F_{\psi} = \frac{1}{d_{A}+1}\Big( P_{\text{EPR}} +\frac{1}{d_{A}}\Big)
\end{align}
for the ideal, noise-free case.

\subsection{Physical interpretation of EPR projection}

\BY{}
Interestingly, Eqn.~\eqref{eq:pfotocstates} suggests that all of the accessible  information about OTOCs probed in a state decoding experiment  are averaged over operators $O_{D}$ on subsystem $D$. 
The physical intuition, as well as the operational interpretation of taking this average  is as follows. 
In classical physics, chaos refers to the sensitive dependence of the system's dynamics on the initial conditions. 
In particular, one can imagine preparing two identical objects, adding a small perturbation to one of them, and then letting them  evolve  under the same Hamiltonian. 
If the system is chaotic,  the outcomes will be drastically different, since a small initial perturbation has an exponentially growing effect. 

In quantum systems, chaos can be probed by preparing a pair of objects with macroscopic entanglement, i.e.~in an EPR pair (or the thermofield double state at finite temperature).
Once again, one can imagine adding a small perturbation to one of the objects, and then letting them evolve under two Hamiltonians, $H$ and $H^{*}$, forward and backward in time, respectively. Without the perturbation, the system remains in an EPR pair since,
\begin{align}
e^{-iHt} \otimes e^{iH^{*}t} |\text{EPR}\rangle =  |\text{EPR}\rangle.
\end{align}
%
However, with a perturbation, the resulting dynamics can cause the system to deviate.
In the above, our perturbation is $O_{A}$ and we are interested in studying how the initial $|\text{EPR}\rangle$ entanglement changes in time.
Operationally, we use the EPR projector, $\Pi_{DD'}$, to check if the entanglement (as seen from $DD'$) has been disturbed by the perturbation or not.
%
%
%
%
%
%
%
This now provides a natural bridge to explain the averaging over operators $O_{D}$ seen in Eqn.~\eqref{eq:pfotocstates}, since $\int d O_{D}\  O_{D}\otimes O_{D}^{*} = \Pi_{DD'}$.

In addition to verifying entanglement, the projector $\Pi_{DD'}$ also has the intriguing effect of ``undoing'' the chaotic dynamics. 
To see this, let us assume that the time-evolution operator $U$ is strongly scrambling so that $\langle \overline{\OTOC} \rangle$ is close to its theoretical minimum $\sim \frac{1}{d_{A}^2}$. By postselecting on $|\text{EPR}\rangle_{DD'}$, one obtains an output state $|\Psi_{\text{out}}\rangle$ [Eqn.~\eqref{psiout}].  Since $\langle \Psi_{\text{in}} |\Pi_{RR'} \Pi_{CC'}\Pi_{DD'} | \Psi_{\text{in}} \rangle = \frac{1}{d_{A}^2}$, one has
\begin{align}
\langle \Psi_{\text{out}} |\Pi_{RR'} \Pi_{CC'}\Pi_{DD'} | \Psi_{\text{out}} \rangle = \frac{1}{d_{A}^2 \langle \overline{\OTOC} \rangle } \approx 1.
\end{align}
Thus, the projector $\Pi_{DD'}$ not only distills an EPR pair on $RR'$, but also undoes the chaotic time-evolution associated with $U$, returning the entire system to a set of EPR pairs!
In particular, if one prepares a quantum state $|\psi\rangle$ on $A$, then the output state will be close to $|\text{EPR}\rangle_{CC'}|\text{EPR}\rangle_{DD'}|\psi\rangle_{R'}$. 
%

%

The fact that the projector $\Pi_{DD'}$ can halt the chaotic dynamics of $U$ is consistent with the traversable wormhole interpretation of the Hayden-Preskill thought experiment \cite{Hayden07,Traversable2017,Yoshida:2017aa}. Indeed, it has been found that the growth of the wormhole interior can be stopped or slowed down by applying certain interactions, and here, $\Pi_{DD'}$ plays the role of resetting the growth of the wormhole. 
Most importantly, this observation provides an additional verification method for our  teleportation-based decoding protocol. Once one measures an EPR pair on $DD'$, it is very likely that one will measure EPR pairs on other pairs of qubits if the experimental procedures are perfect and there is no decoherence.

\section{Teleportation-based Decoding Protocol: Arbitrary noise and decoherence}


In the previous section, we saw that in the \emph{absence} of decoherence, both $P_{\text{EPR}}$ and $F_{\text{EPR}}$ provide the same  information, namely, the value of the averaged OTOC, which in the ideal case, precisely captures the scrambling behavior of the unitary.  
We now turn to our piece de resistance, an analysis of the decoding protocol in the presence of arbitrary noise and imperfections, as characterized via a generic  
quantum channel $\mathcal{Q}$.
The intuition behind the protocol's ability to distinguish between scrambling and decoherence is the redundancy provided by the pair of measurements, $P_{\text{EPR}}$ and $F_{\text{EPR}}$,  in inferring the scrambling behavior of the unitary.  

The protocol proceeds in exactly the same fashion as in the previous section, except that   $\mathcal{Q}$ and $\mathcal{Q}^{*}$ are now applied (rather than $U$ and $U^{*}$). 
A straightforward graphical calculation then yields the probability, $P_{\text{EPR}}$, associated with $\Pi_{DD'}$ as:

\begin{equation}
\begin{split}
&P_{\text{EPR}} =  \langle \widetilde{\OTOC} \rangle \\
&=   \figbox{0.85}{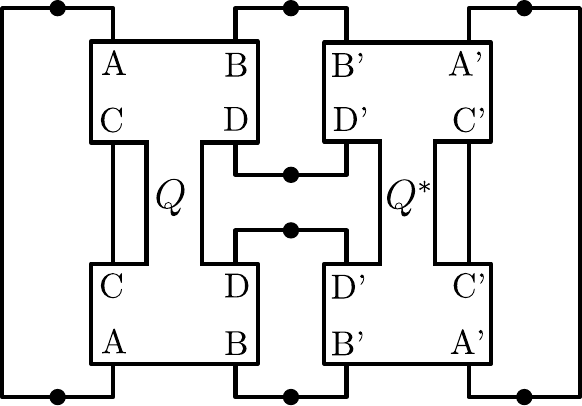} \label{eq-P-channel}\ ,
\end{split}
\end{equation}
As one might recall [Eqn.~\eqref{eq:OTOCtilde_entropy}], $\langle \widetilde{\OTOC} \rangle$ is directly related to the values of the R\'{e}nyi-$2$ entropies, $S_{B'D}^{(2)} + S_{D}^{(2)} - S_{B'}^{(2)}$, meaning that it contains effects from both decoherence and scrambling. 

To measure the mutual information, $I^{(2)}(R,B'D)$, which encodes the true scrambling behavior of the channel $\mathcal{Q}$,
we return to our previous equation for $P_{\text{EPR}}F_{\text{EPR}}$, wherein one finds:
\begin{equation}
\begin{split}
P_{\text{EPR}}F_{\text{EPR}} &=\langle \Psi_{\text{in}}| \Pi_{RR'}\Pi_{DD'}\otimes I_{CC'} | \Psi_{\text{in}}\rangle \\
&= \frac{1}{d_{A}^2}\ \figbox{0.9}{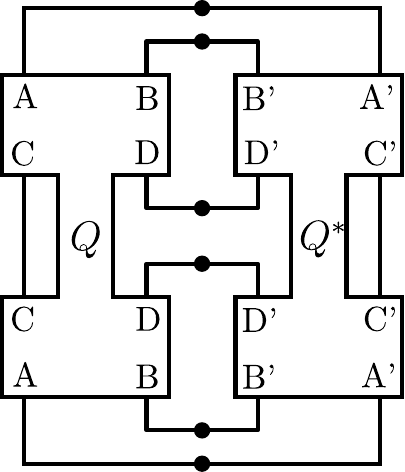}  \\ &= \frac{d_{C} \Tr(\rho_{RB'D}^2) }{d_{A}^2} =  \frac{ \delta }{d_{A}^2},
\end{split}
\label{eq:pfdelta}
\end{equation}
where  $\delta \equiv \frac{2^{I^{(2)}(R,B'D)}}{2^{S_{B'D}^{(2)} + S_{D}^{(2)} - S_{B'}^{(2)}}}= 2^{I^{(2)}(R,B'D)}\times P_{\text{EPR}}$ is precisely our previously defined noise parameter [Eqn.~\eqref{eq:deltanoiseparam}]!
We emphasize that equations \eqref{eq-P-channel} and \eqref{eq:pfdelta} are precisely the ``noisy-quantum-channel'' analogs of equations \eqref{eq-P} and \eqref{pf-nonoise} for the noise-free case.
The decoding fidelity after post-selection is then given by:
\begin{align}
F_{\text{EPR}} = \frac{ 2^{I^{(2)}(R,B'D) } }{ d_{A}^2}.
\end{align}
Thus, the success of teleportation implies true scrambling (i.e.~large $I^{(2)}(R,B'D)$) for a generic quantum channel,  $\mathcal{Q}$. 
Moreover, by measuring both $P_{\text{EPR}}$ and $F_{\text{EPR}}$, one can directly compute $\delta$, thereby characterizing the amount of noise in the quantum channel.

To see this in action, let us now return to the case where $\mathcal{Q}$ reflects a depolarizing channel  [Eqn.~\eqref{eq:noise}].
 In this situation, the measurement of $\delta$ via $P_{\text{EPR}}$ and $F_{\text{EPR}}$ immediately provides insight into the amount of dissipation in the system (given by probability $p$), since
\begin{align}
\delta = 
\Big[  (1-p)^2  + (2p -p^2)\frac{1}{d_{D}^2} \Big]. 
\end{align}
While experimental decoherence cannot always be recast simply as depolarization, this expression serves as an operational (and quantitative) measure of extrinsic experimental noise.

For the case of state decoding, an analogous calculation reveals that the error parameter $\delta$ is given by:
\begin{align}
\int d\psi\ P_{\psi}F_{\psi} = \frac{1}{d_{A}+1}\Big( P_{\text{EPR}} +\frac{\delta}{d_{A}}\Big).
\end{align}
Interestingly, we note that as an alternative strategy, one can also study the effect of decoherence for a specific input state by observing possible violations of the bound in Eqn.~\eqref{eq:bound-state}.


\section{Teleportation-based Decoding Protocol: Coherent Errors}

\subsection{Distinguishing scrambling from coherent errors}

In the previous section, we focused on the case of a generic noisy quantum channel and more specifically, on the effects of depolarization. 
In this subsection, motivated by recent experiments \cite{garttner2017measuring,li2017measuring}, we will consider the case of coherent unitary errors (i.e.~systematic over or under-rotations), which lead to imperfect ``backwards'' time-evolution (but no non-unitary decoherence). 
In particular, we will investigate the situation where the time-evolution operator is given by $U\otimes V^{*}$ (rather than $U\otimes U^{*}$, which we assume to be the desired ideal case). For simplicity, let us assume that all other operations, including the initial preparation of EPR pairs and the final readout measurements are error-free \footnote{Note that coherent errors in the initial EPR preparation can also be absorbed into the definition of $V$.}. 

In this scenario, the probability of measuring  $|\text{EPR}\rangle_{DD'}$ is given  by:
\begin{align}
P_{\text{EPR}} = \iint d O_{A}dO_{D} \langle O_{A}O_{D}(t) O_{A}^{\dagger}O^{\dagger}_{D_{V}}(t)\rangle = \nonumber \\ \figbox{0.9}{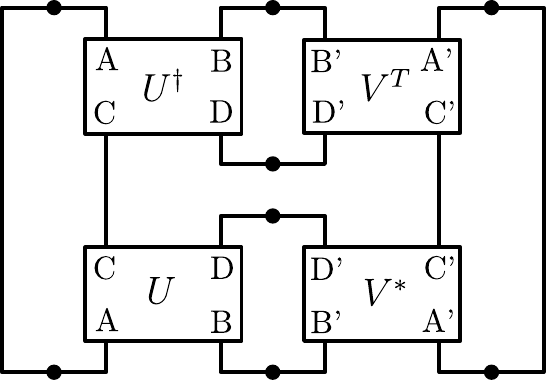}
\label{pcoherenterrors}
\end{align}
where $O_{D}(t)=UO_{D}U^{\dagger}$ and $O_{D_V}(t)=VO_{D}V^{\dagger}$ are time-evolved by different unitaries, $U$ and $V$, respectively. 
%
%
A simple graphical calculation yields the product, $P_{\text{EPR}}F_{\text{EPR}}$, as
\begin{align}
P_{\text{EPR}}F_{\text{EPR}} = \frac{1}{d_{A}^2}\ \figbox{1.0}{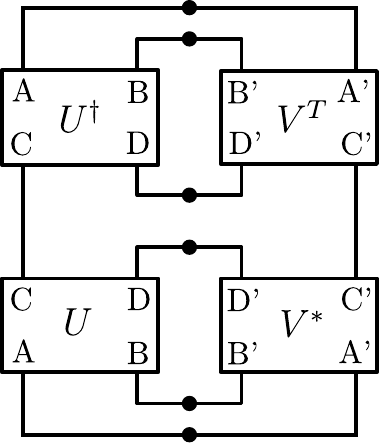} \ \leq \frac{1}{d_{A}^2},
\label{pfcoherenterrors}
\end{align}
which is strictly smaller than the ideal case (i.e. when $U=V$), where $P_{\text{EPR}}F_{\text{EPR}}=\frac{1}{d_{A}^2}$.  Again, we emphasize that equations \eqref{pcoherenterrors} and \eqref{pfcoherenterrors} are precisely the ``coherent-error'' analogs of equations \eqref{eq-P} and \eqref{pf-nonoise} in the ideal case.

By analogy to Eqn.~\eqref{eq:pfdelta}, this suggests that one can define a noise parameter, $\eta$, for \emph{coherent} errors as follows
\begin{align}
P_{\text{EPR}}F_{\text{EPR}} = \frac{\eta}{d_{A}^2}.
\end{align}
While $\eta$ and $\delta$ effectively measure the same diagram, $\eta$ cannot be interpreted in terms of entropy since $U\otimes U^{*}$ is performed incorrectly. 

Moreover, the physical interpretation of $\eta$ is quite different from that of $\delta$, which characterizes the strength of decoherence. 
In particular, we note that a natural measure of the amount of coherent error is provided by the composite unitary operator, $E = U^\dagger V$.  In the error-free, ideal case, $E$ simply corresponds to the identity operation. The noise parameter, $\eta$, is related to $E$ as follows
\begin{align}
\eta &= \ \figbox{1.0}{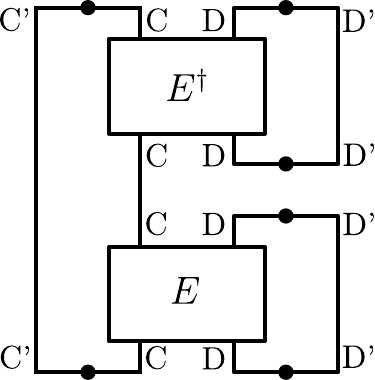} \nonumber \\ &=\text{Tr}\left(\mathbb{I}\otimes \Pi_{DD'}E\Big( \frac{\mathbb{I}}{d_{C}}\otimes \Pi_{DD'} \Big) E^{\dagger} \right).
\label{eq:etaerror}
\end{align}
The right hand side of Eqn.~\eqref{eq:etaerror} is  the $2$-norm overlap between $E|_{D}$ and $\mathbb{I}_{D}$; here, $E|_{D}$ is the quantum channel defined on $D$ by tracing out the degrees of freedom on $C$. If one takes $D$ to be the whole system, then $\eta$ is simply the $2$-norm overlap between $E$ and  $\mathbb{I}$, precisely capturing the amount of deviation between the composite unitary, $U^{\dagger}V$, and the identity.
Finally, we note that $0\leq \eta  \leq 1$, where the lower bound is saturated when $E= \mathbb{I}\otimes O_{D} $ for any traceless operator $O_{D}$. This contrasts with the decoherence noise parameter, $\delta$, which is lower bounded by $\min(\frac{1}{d_{A}^2},\frac{1}{d_{D}^2})$.

\subsection{Bounding the OTOC via $\eta$ in the case of coherent errors}

Intriguingly, under certain physical assumptions, one can utilize the measured value of $\eta$ to upper bound the true value of the OTOC that would have been measured without coherent errors (i.e.~if $U=V$). 
In particular, we would like to compare the following two quantities:
\begin{align}
\langle  O_{A}O_{D}(t)O_{A}^{\dagger}O_{D}^{\dagger}(t)\rangle \qquad
\langle  O_{A}O_{D}(t)O_{A}^{\dagger}O_{D_V}^{\dagger}(t)\rangle,
\label{eq:twootocseqn}
\end{align}
and use the second, which is measured via $P_{\text{EPR}}$, to bound the first. For simplicity, let us assume that $O_{A},O_{D}$ are Pauli operators. 

We will also make and justify a second assumption. In particular, consider an OTOC of the form, 
$\langle O_{A}(0) O_{D}(t) O_{A}^{\dagger} O_{D}'^{\dagger}(t) \rangle$
where the two operators, $O_{D}$ and $O_{D}'$, are both time-evolved by $U$. 
We will assume that
\begin{align}
\langle O_{A}(0) O_{D}(t) O_{A}^{\dagger} O_{D}'^{\dagger}(t) \rangle  \approx 0    \label{eq:assumption-OTOC}
\end{align}
so long as $\Tr(  O_{D} O_{D}'^{\dagger}) =0$.~The intuition behind this assumption is as follows.~At $t=0$, if regions $A$ and $D$ do not overlap, then $\langle O_{A}(0) O_{D}(t) O_{A}^{\dagger} O_{D}'^{\dagger}(t) \rangle=\Tr(  O_{D} O_{D}'^{\dagger})= 0$.
Then, since OTOCs generically decay under ergodic time-evolution, one expect the above expectation value to remain small \emph{throughout} the time-evolution.

To proceed, it will be useful to define a new un-evolved (e.g.~time $t=0$) operator $O_E = E O_{D}E^{\dagger}$, which corresponds to the conjugation of $O_{D}$ by the composite unitary $E$. The subsequent time evolution of this operator via the unitary $U$ is given by: $UEO_{D}E^{\dagger}U^{\dagger} =   VO_{D}V^{\dagger} = O_{D_V}(t)$. Then, we have 
\begin{align}
\langle  O_{A}O_{D}(t)O_{A}^{\dagger}O_{D_V}^{\dagger}(t)\rangle = \langle  O_{A}O_{D}(t)O_{A}^{\dagger}O_{E}^{\dagger}(t)\rangle \label{eq:OTOC-exp}
\end{align}
where $O_{E}(t)=UO_{E}U^{\dagger}$. 
Let us now expand the composite unitary, $E$, in terms of Pauli operators, $P$ and $Q$:
\begin{align}
E = \sum_{P,Q} \alpha_{P,Q} P \otimes Q, 
\end{align}
where $P$, $Q$ act on subsystems $C$, $D$, respectively and $\sum_{P,Q}|\alpha_{P,Q}|^2 = 1$ \footnote{Note that this normalization condition is implied by the unitarity of $E$.}.  Plugging this into our expression for $\eta$, one obtains
\begin{align}
\eta = \sum_{P} |\alpha_{P,\mathbb{I}}|^2.
\end{align}

Let us also expand $O_{E}$ in terms of Pauli operators,
\begin{align}
O_{E}= \sum_{P,Q} \beta_{P,Q} P \otimes Q,\label{eq:OD-hat}
\end{align}
where again $\sum_{P,Q}|\beta_{P,Q}|^2 = 1$. Plugging this expression back into Eqn.~\eqref{eq:OTOC-exp} yields,
\begin{align}
\langle  O_{A}O_{D}(t)O_{A}^{\dagger}&O_{E}^{\dagger}(t)\rangle = 
\sum_{P,Q} \beta_{P,Q} \langle  O_{A}O_{D}(t)O_{A}^{\dagger}(P\otimes Q)(t)\rangle \nonumber \\
&\approx \beta_{\mathbb{I},O_{D}}\langle  O_{A}O_{D}(t)O_{A}^{\dagger}O_{D}^{\dagger}(t)\rangle,
\label{betadecompotoc}
\end{align}
where we have  used our assumption [Eqn.~\eqref{eq:assumption-OTOC}] to drop all terms with $Q \neq O_D$ in going from the first to second line. 

Noting that $\beta_{\mathbb{I},O_{D}} = \frac{1}{d}\Tr ( O_{D} O_{E}^{\dagger} )$, allows us to bound it as follows:
\begin{align}
&\beta_{\mathbb{I}, O_{D} } 
= \frac{1}{d} \Tr\left[  (\mathbb{I} \otimes O_{D} ) E(\mathbb{I} \otimes O_{D}^{\dagger} )E^{\dagger} \right] \nonumber \\
&= \frac{1}{d} \Tr \left[  \sum_{P,Q} |\alpha_{P,Q}|^2 ( \mathbb{I} \otimes O_{D} ) (P \otimes Q )( \mathbb{I} \otimes O_{D}^{\dagger} ) (P \otimes Q ) \right]  \nonumber \\
&= \sum_{P} |\alpha_{P,\mathbb{I}}|^{2} + \sum_{P}\sum_{Q\not=\mathbb{I}} \pm|\alpha_{P,Q}|^{2}  \geq 2\eta -1. \label{eq:bound-beta}
\end{align}
Here,  the $\pm$-signs in the final line correspond to the case where $O_{D}$ and $Q$ commute/anti-commute, respectively.
Thus, the lower bound corresponds to the case where all non-zero $\alpha_{P,Q}$ come with a negative sign.



Finally, combining Eqns.~\eqref{eq:OTOC-exp}, \eqref{betadecompotoc}, and \eqref{eq:bound-beta} yields the following bound:
\begin{align}
\langle  O_{A}O_{D}(t)O_{A}^{\dagger}&O_{D_V}^\dagger(t)\rangle   =\langle  O_{A}O_{D}(t)O_{A}^{\dagger}O_{E}^\dagger(t)\rangle \nonumber \\
&= \beta_{\mathbb{I},O_{D}}\langle  O_{A}O_{D}(t)O_{A}^{\dagger}O_{D}^{\dagger}(t)\rangle \nonumber \\
&\geq ( 2\eta - 1) \langle  O_{A}O_{D}(t)O_{A}^{\dagger}O_{D}^{\dagger}(t)\rangle.
\end{align}
Thus, in the case of coherent errors corresponding to imperfect backwards time evolution, the experimentally \emph{measured} value of the averaged OTOC (via for example $P_\text{EPR}$) explicitly bounds the actual ideal OTOC:
\begin{align}
\langle \overline{\text{OTOC}} \rangle \le \frac{P_{\text{EPR}}}{ 2\eta -1 }.
\end{align}
We note that this bound is only valid for $\eta > 0.5$.

Two additional remarks. First, it is worth pointing out that the value of $\beta_{\mathbb{I}, O_{D} }$ can be  directly measured via $\langle  O_{A}O_{D}(t)O_{A}^{\dagger}O_{D_V}^{\dagger}(t)\rangle$ with $O_{A}=\mathbb{I}$, since $\beta_{\mathbb{I}, O_{D} } = \langle O_{D}O_{E}^{\dagger}(t)\rangle=\langle O_{D}(t) O_{D_V}^{\dagger}(t)\rangle$.
Second, in a generic chaotic system, one expects the $\pm$-signs in Eqn.~\eqref{eq:bound-beta} to appear randomly. Under this assumption, one can make the following approximation: 
\begin{align}
\beta_{\mathbb{I}, O_{D} }  \approx \sum_{P} |\alpha_{P,\mathbb{I}}|^2 = \eta,
\end{align}
which enables us to obtain an estimate for the actual \emph{value} of the OTOC and not simply a bound,
\begin{align}
 \langle  O_{A}O_{D}(t)O_{A}^{\dagger}O_{D}^{\dagger}(t)\rangle \approx \frac{1}{\eta}\langle  O_{A}O_{D}(t)&O_{A}^{\dagger}O_{D_V}^\dagger(t)\rangle.
\end{align}


\section{Bounding the Mutual Information  via the decoding fidelity}

\subsection{Mutual Information Bound}

In the previous section, we have shown that in the case of coherent errors, one can utilize $\eta$ as extracted from $P_{\text{EPR}}$ and $F_{\text{EPR}}$ to formally bound the true (i.e.~error-free) value of the averaged OTOC. However, this proof explicitly hinges on the unitarity of the composite channel $E$ and is thus inapplicable to the generic situation with decoherence. Moreover, in the presence of decoherence, it becomes ambiguous to define what precisely the \emph{value} of the OTOC is \footnote{The intuition behind this ambiguity is that for a generic quantum channel, one can decompose its action using Kraus operators but this decomposition is not unique \cite{nielsen2010quantum}.}; rather, as we have previously seen, a better characterization for quantum scrambling is provided by the mutual information. 

To this end, in this section, we demonstrate that for \emph{arbitrary} quantum channels, one can derive a bound
on the mutual information, $I^{(2)}(R,B'D)$,  using only the decoding fidelity, $F_{\text{EPR}}$ \footnote{The key point here is that  $F_{\text{EPR}}$ always provides a lower bound on the \emph{mutual information} between $R$ and $B'D$, regardless of the nature of  experimental errors. Stated differently, in the context of the black hole information problem, the fact that one can retrieve a quantum state from the Hawking radiation (i.e.~the teleportation is successful) implies that the system has scrambled, regardless of how one performs the decoding.}. When applied to the case of purely unitary errors (i.e.~the previous section), this leads to a somewhat weaker bound on $\langle \overline{\text{OTOC}} \rangle$. 
%

\begin{figure}
\centering\includegraphics[width=1.5in]{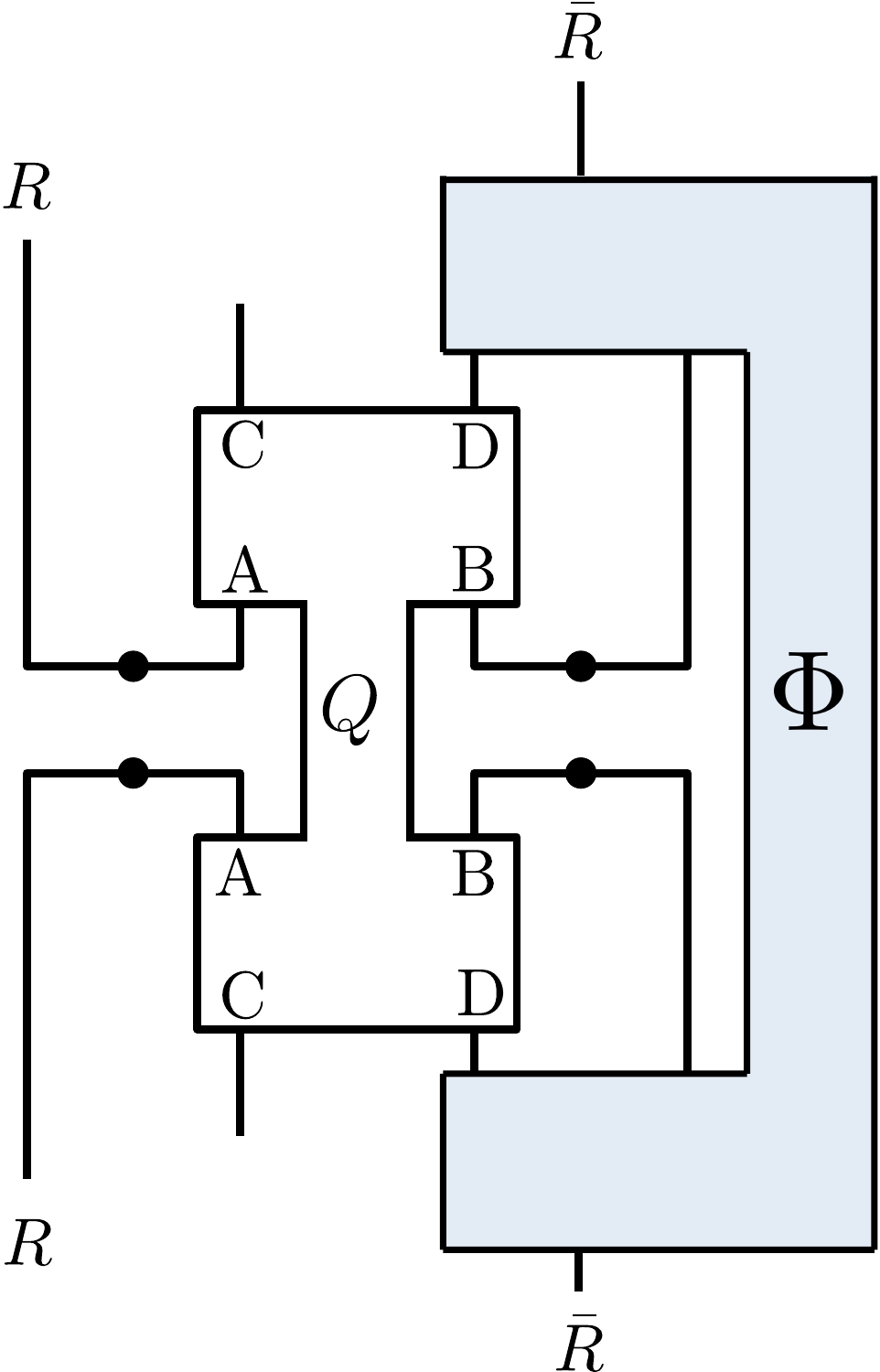}
\caption{Schematic representation of the decoding protocol as the distillation of an EPR pair on $R\bar{R}$. The initial state is $|\text{EPR}\rangle_{RA}|\text{EPR}\rangle_{BB'}$. The quantum channel $\mathcal{Q}$ describes the time-evolution of the system and $\Phi$ represents an arbitrary decoding operation acting only on $B'D$.
}
\label{fig-general}
\end{figure}

%
To treat experimental imperfections on a fully general footing, we consider time-evolution via the quantum channel $\mathcal{Q}$ and an arbitrary decoding operation $\Phi$, acting non-trivially only on $B'D$ (Fig.~\ref{fig-general}). As previously discussed, the goal of this decoding operation is to distill an EPR pair on $R\bar{R}$, where $\bar{R}$ represents a subset of the qubits in $B'D$ with the same dimension as $R$ (e.g. $|R|=|\bar{R}|$).
Let us assume that $\Phi$, an arbitrary completely-positive trace-preserving map, outputs a normalized state supported on $R\bar{R}$:
\begin{align}
\Phi: \rho_{RB'D} \rightarrow \sigma_{R\bar{R}}.
\end{align}
Since the decoding operation acts locally on $B'D$, it \emph{cannot} increase entanglement between $R$ and $B'D$, i.e.~the mutual information satisfies $I(R,B'D)\geq I(R,\bar{R})$ \footnote{Note that we are using the von Neumann mutual information here.}.
Since $I(R,\bar{R})$ can be lower bounded via $F_{\text{EPR}}$, any non-trivial decoding fidelity always signifies quantum scrambling even in the presence of arbitrary imperfections.

Deriving a rigorous lower bound on either the von Neumann  or R\'{e}nyi-$2$ mutual informations in terms of $F_{\text{EPR}}$ is tremendously challenging, owing in part,  to the  existence of  fine-tuned adversarial examples that tend to make the bounds loose  in physically relevant situations.
%
%
This is ameliorated by making the assumption that the decoding fidelity is independent of the input state $|\psi\rangle$. 
One expects this assumption to be approximately valid for strongly interacting systems without conserved quantities after the system locally thermalizes. This assumption also excludes the trivial decoding protocol which returns a fixed state regardless of  input state.


To begin, we note that  $\rho_{R}$ is a maximally mixed state [Eqn.~(11)], implying that $\sigma_{R}$ is also maximally mixed, since $\Phi$ acts only locally on $B'D$.
Moreover, our assumption that the decoding fidelity does not depend on the input state implies that $\sigma_{\bar{R}}$ is  a maximally mixed state as well. 
While generally true, this statement is particularly easy to see in the case where $R$ and $\bar{R}$ consist  of only single qubits. 
In particular, we can use our previous trick and decompose $\sigma_{R\bar{R}}$ in terms of Pauli operators: $\sigma_{R\bar{R}} = \sum_{P,Q} \gamma_{P,Q} P\otimes Q$, for $P,Q \in \{\mathbb{I}, X,Y,Z\}$. 
%
%
Since $\sigma_{R}$ is maximally mixed, one has that $\gamma_{P, \mathbb{I}}=0$ for all $P\not=\mathbb{I}$.

For an input state  $|\psi\rangle$, the quantum state on $\bar{R}$ is given by
\begin{align}
d_{R}(|\psi\rangle\langle \psi| \otimes \mathbb{I}_{\bar{R}}) \sigma_{R\bar{R}}(|\psi\rangle\langle \psi| \otimes \mathbb{I}_{\bar{R}}) 
\end{align}
where $d_{R}$ is a normalization constant. The decoding fidelity can then be written as 
\begin{align}
F_{\psi}= d_{R}\Tr ( |\psi\rangle\langle \psi| \otimes  |\psi^{*}\rangle\langle \psi^{*}| \sigma_{R\bar{R}}  ).
\end{align}
Noting that $|0\rangle\langle 0 | = \frac{I+Z}{2}$ and  $|1\rangle\langle 1 | = \frac{I-Z}{2}$, implies $\gamma_{Z, \mathbb{I}}+\gamma_{\mathbb{I},Z}=0$;  since $\gamma_{Z,\mathbb{I}}=0$, one also has that
$\gamma_{\mathbb{I}, Z}$=0. The same analysis leads to $\gamma_{\mathbb{I},P}=0$ for all non-identity Pauli operators $P$. Thus, $\sigma_{\bar{R}}$ is also a maximally mixed state.

Having shown that $\sigma_{\bar{R}}$ is a maximally mixed state, we are now ready to lower bound the mutual information. The EPR projector and the decoding fidelity  of the distilled quantum state, $\sigma_{R\bar{R}}$, are given by: $\Pi_{R\bar{R}}= |\text{EPR}\rangle\langle \text{EPR}|_{R\bar{R}}$ and $F_{\text{EPR}} = \Tr( \Pi_{R\bar{R}} \rho_{R\bar{R}} )$, respectively.
Then using the Cauchy-Schwartz inequality,  one immediately arrives at the following bound:
\begin{align}
&S^{(2)}_{R\bar{R}}= - \log_{2} \Tr( \rho_{R\bar{R}}^2 ) \leq \nonumber \\ &- \log_{2} \Tr( \Pi_{R\bar{R}}\rho_{R\bar{R}}) \Tr( \Pi_{R\bar{R}}\rho_{R\bar{R}}) =  -  2\log_{2}F_{\text{EPR}},
\end{align}
implying that the mutual information satisfies:
\begin{align}
I^{(2)}(R,\bar{R}) = S_{R} + S_{\bar{R}} - S_{R\bar{R}} \ge  2 \log_{2} d_{R} + 2 \log_{2} F_{\text{EPR}}.
\end{align}
In order to utilize the monotonicity of mutual information \cite{nielsen2010quantum}, we will make the additional technical assumption that the  R\'{e}nyi-$2$ and von Neumann entropies are close to one another \footnote{Recall that R\'{e}nyi-$2$ mutual information is not monotonically decreasing in general \cite{suppinfo}.}. 
This then leads to our final result, lower bounding the mutual information in terms of the decoding fidelity:
\begin{align}
I(R,B'D)\approx I^{(2)}(R,B'D) \ge  2 \log_{2} d_{R} + 2\log_{2} F_{\text{EPR}}.
\end{align}

%


\subsection{OTOC Bound for Coherent Errors}

While the previous subsection focused on the  case of \emph{arbitrary} quantum channels, one can also apply the derived bound to the situation where only coherent errors are present.
To this end,  let us return to scenario described in Sec.~IVb, where the time-evolution is given by $U\otimes V^{*}$.
As we have already seen, the measurement of $P_{\text{EPR}}$ corresponds to 
\begin{align}
P_{\text{EPR}} =  \int dO_{A} d O_{D} \langle O_{A}O_{D}(t)O_{A}^{\dagger}O_{D_V}^{\dagger}(t) \rangle,
\end{align}
which includes the effect of unitary errors associated with $E = U^\dagger V \neq \mathbb{I}$. In analogy to Sec.~IVc, the true OTOC, which would have been measured if the experiment did not contain such unitary errors is given by: 
\begin{align}
\overline{\langle \OTOC \rangle} = \int dO_{A} d O_{D} \langle O_{A}O_{D}(t)O_{A}^{\dagger}O_{D}^{\dagger}(t) \rangle.
\end{align}
Since $ \overline{\langle \OTOC \rangle}= 2^{-I^{(2)}(A,BD)}$, our above bound on the mutual information also immediately bounds $ \overline{\langle \OTOC \rangle}$ in the case of purely coherent errors:
\begin{align}
\overline{\langle \OTOC \rangle} \leq \frac{1}{d_{R}^2 F_{\text{EPR}}^2}. 
\end{align}

\section{Experimental Implementation}

Having detailed a teleportation protocol that explicitly enables experiments to distinguish between decoherence and quantum information scrambling \footnote{Notable examples of fast quantum information scramblers include: the SYK model~\cite{Kitaev:2014t2},  $k$-local random spin models~\cite{Erdos:2014aa} and  random quantum circuits~\cite{Dankert09}. }, we now propose two specific examples of scrambling Clifford circuits \cite{suppinfo} amenable to near-term experiments in small-scale quantum simulators \cite{zhang2017observation,bernien2017probing}. 
%
%
%

\subsection{Qubit Clifford Scrambler} 

Let us consider the following $3$-qubit unitary operator:
\begin{align}
U = \figbox{0.9}{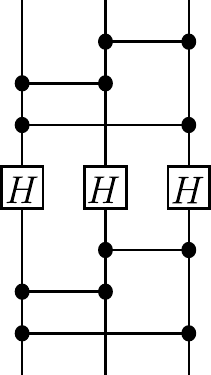}
\end{align}
where $H$ represents a Hadamard gate, while two-qubit, control-$Z$ gates ($|i,j\rangle \rightarrow (-1)^{ij}|i,j\rangle$) are depicted as horizontal lines (ending in dots). 
This unitary is  maximally scrambling since all one-body Pauli operators are delocalized into three-body Pauli operators under $U$ \cite{Pastawski15b}. From the perspective of decoding, this delocalization implies that Bob can collect any pair of qubits (from among the three possible pairs in Figure 3) and perform a projective measurement in order to decode Alice's state. 
To be concrete, the full decoding protocol is illustrated in Figure 3. 
\begin{figure}
\centering\includegraphics[width=2.3in]{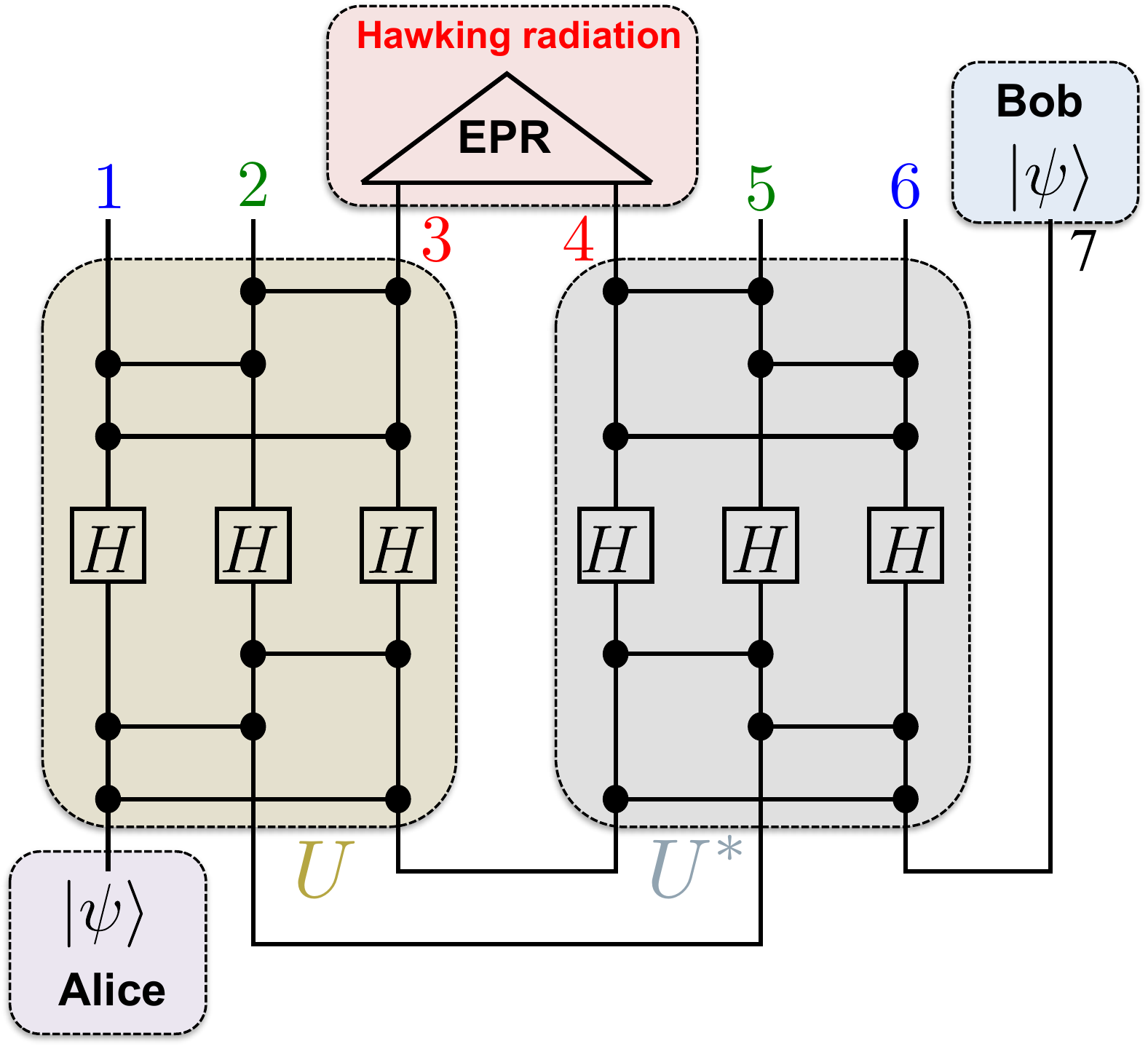}
\caption{Decoding circuit based upon a 3-qubit Clifford scrambler. Alice's quantum state $|\psi\rangle$ is supported on qubit-$1$, while Bob's quantum register corresponds to qubit-$7$. 
The left (beige) and right (gray) Hilbert spaces have the following correspondences $1 \leftrightarrow 6$, $2 \leftrightarrow 5$ and $3 \leftrightarrow 4$ (with respect to $U\otimes U^{*}$).
By performing an EPR projection on qubits $3$ and $4$, Bob teleports Alice's quantum state to his register qubit. In the case of this Clifford scrambler, Bob could also have achieved teleportation by performing EPR projections on either qubits $\{1,6\}$ or $\{2,5\}$. This distinguishes the Clifford scrambler from other more trivial (non-scrambling) unitaries (i.e.~a SWAP gate), where teleportation only occurs for EPR projection on a specific pair of qubits. 
}
\label{fig-qubitexpt}
\end{figure}

Two comments are in order. 
In particular, for a Haar random unitary, one expects $\langle \overline{\OTOC}\rangle_{\text{S}} = \frac{7}{16}$, whereas our circuit exhibits: $\langle \overline{\OTOC}\rangle_{\text{S}} = \frac{1}{4}$.
This discrepancy arises from finite size effects, since one expects a Haar random unitary to  saturate the lower bound of $1/4$ only in the limit of large systems, i.e.~$d,d_{D}\rightarrow \infty$ while fixing $d_{R}=2$.
On the other hand, our Clifford circuit saturates this lower bound by construction but has certain non-generic features \cite{suppinfo}.
Second, as we briefly alluded to in Sec.~IIIc, it is also possible to explore circuits that scramble only classical information:
\begin{align}
U = \figbox{1.0}{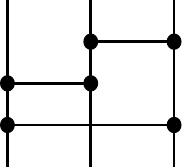}.
\end{align}
In this case, teleportation  occurs only for computational basis states.

\subsection{Qutrit Clifford Scrambler} 

While we presented the minimum case of interest for qubits in the previous subsection, by increasing the on-site Hilbert space, one can realize slightly more complex circuits in even smaller systems. To this end, motivated by the advent of physical qutrit implementations ranging from solid-state spin defects and superconducting circuits to orbital angular momentum states of photons, we describe a simple qutrit Clifford scrambler.

To begin, we denote a qutrit as a  three-state quantum spin with basis: $|0\rangle, |1\rangle, |2\rangle$.  An elementary entangling gate between two qutrits can be achieved via the following controlled-NOT gate:
\begin{align}
\text{CNOT}_{1\rightarrow 2}|i,j\rangle = |i,i+j\rangle \qquad \text{modulo $3$}
\end{align}
where the subscript $1\rightarrow 2$ indicates that the control is qutrit-$1$ and the target is qutrit-$2$. Switching the control and target realizes an analogous operation: $\text{CNOT}_{2\rightarrow 1}|i,j\rangle = |i+j,i\rangle
 \hspace{1mm}\text{mod $3$}$.

Let us now consider the following qutrit unitary:
\begin{align}
U = \text{CNOT}_{2\rightarrow 1}\text{CNOT}_{1\rightarrow 2},
\end{align}
which can be explicitly decomposed as $U|i,j\rangle = |2i+j,i+j\rangle$ or graphically re-expressed as:
\begin{align}
U = \figbox{1.0}{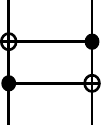}.
\label{eq:qutritunitarycnots}
\end{align}
To understand the scrambling properties of this unitary, we will explore how the qutrit Pauli operators are transformed under the circuit; in particular, let us consider the following qutrit Pauli's: $X = \sum_{j=0}^{2} |j+1\rangle\langle j|$ and $Z = \sum_{j=0}^2 \omega^j |j\rangle\langle j|$ where $\omega = e^{i\frac{2\pi}{3}}$. 

One finds that these operators are transformed as follows:
\begin{align}
&U(Z\otimes I )U^{\dagger} = Z\otimes Z^2 \nonumber \\ &U(I\otimes Z )U^{\dagger} = Z^2\otimes Z^2 \nonumber \\
&U(X\otimes I )U^{\dagger} = X^2\otimes X  \nonumber \\ &U(I\otimes X )U^{\dagger} = X\otimes X.
\end{align}
Thus, as in the qubit case, we observe that the unitary transforms any non-identity one-body Pauli operator  into a two-body operator. 
This property is essential for the delocalization of quantum information and enables the construction of a similar decoding protocol:
\begin{align}
\figbox{1.0}{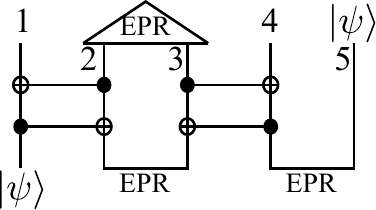}.
\end{align}
By performing an EPR projection on either qutrits $\{2,3\}$ or $\{1,4\}$, Bob successfully teleports Alice's quantum state from qutrit-$1$ to qutrit-$5$.

\subsection{Distinction from conventional quantum teleportation} 

The importance of being able to perform teleportation by projecting \emph{either} pair of qutrits (or in the previous case, any of the three qubit pairs) is most easily seen by considering the effect of a SWAP gate, $\text{SWAP}|i,j\rangle = |j,i\rangle$, or graphically:
\begin{align}
\textrm{SWAP} = \figbox{1.0}{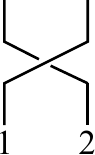}.
\end{align}
From the perspective of scrambling, a SWAP gate is totally trivial since it does not generate any entanglement; thus, its decoding behavior must be markedly different from that of the maximally scrambling $U$ in Eqn.~\eqref{eq:qutritunitarycnots}.

Replacing $U$ with the SWAP gate in the decoding protocol leads to the following: 
\begin{align}
\figbox{1.0}{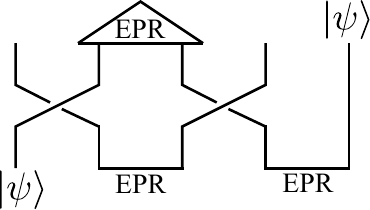},
\end{align}
which is simply ordinary quantum teleportation \cite{nielsen2010quantum, bouwmeester1997experimental,furusawa1998unconditional}. Crucially, this teleportation \emph{only} works when Bob projects on qutrits $\{2, 3\}$ and fails if he attempts to  project on qutrits $\{1, 4\}$. Herein lies the essential feature of a maximally scrambling unitary: Successful decoding and teleportation occur regardless of which pair of qutrits (or qubits) one chooses to collect and project, precisely indicating the full delocalization of quantum information across the circuit. 

\section{Conclusion}

We have demonstrated that one can distinguish between scrambling and decoherence in strongly interacting quantum systems by utilizing a teleportation-based decoding protocol. Our protocol enables the explicit extraction of a ``noise parameter'', which can be used to bound the value of out-of-time-ordered correlation functions in the case of coherent errors. More generally, even for arbitrary imperfections, the teleportation fidelity acts as a metric 
for quantum scrambling and enables the bounding of the mutual information between subsystems. Thus, our protocol represents the first example of an experimental method, which can unambiguously characterize the 
delocalization of quantum information within a system's own degrees of freedom and differentiate this from entanglement with an extrinsic environment.

Our work opens the door to a number of intriguing future directions. 
First,  by systematically exploring the state-dependent decoding fidelity in the presence of different forms of decoherence, one may be able to study the transition from  classical to quantum chaos. 
Second, in this work, we have mainly focused on decoherence as an adversary to quantum scrambling. 
However, the pronounced sensitivity of scrambling dynamics to the presence of decoherence suggests that one may be able to utilize our protocol as a particularly efficient ``noise'' spectroscopy tool. 
Finally, an interesting question that has received much recent attention, and which goes under the moniker of quantum supremacy, is whether quantum devices without error correction can perform computational tasks beyond the capabilities of classical computers \cite{preskill2018quantum}.
 It has been suggested that the  simulation of random quantum  circuits may be an ideal platform for this purpose~\cite{boixo2016characterizing}. 
 Since OTOCs are natural probes of pseudorandomness, it may  be possible to generalize our protocol to explore such questions. 

\emph{Acknowledgements---}We gratefully acknowledge the insights of and discussions with P. Hess, A. Kyprianidis, G. Pagano, J. Zhang, V. Ramasesh, M. Blok, I. Siddiqi, C. Monroe, and Q. Zhuang.
We are particularly indebted to T. Schuster for a careful reading of the manuscript and many helpful discussions.
This work was supported by the DOE under contract PH-COMPHEP-KA24 and the Office of Advanced Scientific Computing Research, Quantum Algorithm Teams Program.


\mciteSetMidEndSepPunct{}{\ifmciteBstWouldAddEndPunct.\else\fi}{\relax}
%
\bibliography{reference2018Jan}{}

\begin{thebibliography}{71}%
\makeatletter
\providecommand \@ifxundefined [1]{%
 \@ifx{#1\undefined}
}%
\providecommand \@ifnum [1]{%
 \ifnum #1\expandafter \@firstoftwo
 \else \expandafter \@secondoftwo
 \fi
}%
\providecommand \@ifx [1]{%
 \ifx #1\expandafter \@firstoftwo
 \else \expandafter \@secondoftwo
 \fi
}%
\providecommand \natexlab [1]{#1}%
\providecommand \enquote  [1]{``#1''}%
\providecommand \bibnamefont  [1]{#1}%
\providecommand \bibfnamefont [1]{#1}%
\providecommand \citenamefont [1]{#1}%
\providecommand \href@noop [0]{\@secondoftwo}%
\providecommand \href [0]{\begingroup \@sanitize@url \@href}%
\providecommand \@href[1]{\@@startlink{#1}\@@href}%
\providecommand \@@href[1]{\endgroup#1\@@endlink}%
\providecommand \@sanitize@url [0]{\catcode `\\12\catcode `\$12\catcode
  `\&12\catcode `\#12\catcode `\^12\catcode `\_12\catcode `\%12\relax}%
\providecommand \@@startlink[1]{}%
\providecommand \@@endlink[0]{}%
\providecommand \url  [0]{\begingroup\@sanitize@url \@url }%
\providecommand \@url [1]{\endgroup\@href {#1}{\urlprefix }}%
\providecommand \urlprefix  [0]{URL }%
\providecommand \Eprint [0]{\href }%
\providecommand \doibase [0]{http://dx.doi.org/}%
\providecommand \selectlanguage [0]{\@gobble}%
\providecommand \bibinfo  [0]{\@secondoftwo}%
\providecommand \bibfield  [0]{\@secondoftwo}%
\providecommand \translation [1]{[#1]}%
\providecommand \BibitemOpen [0]{}%
\providecommand \bibitemStop [0]{}%
\providecommand \bibitemNoStop [0]{.\EOS\space}%
\providecommand \EOS [0]{\spacefactor3000\relax}%
\providecommand \BibitemShut  [1]{\csname bibitem#1\endcsname}%
\let\auto@bib@innerbib\@empty
\bibitem [{\citenamefont {Deutsch}(1991)}]{deutsch1991quantum}%
  \BibitemOpen
  \bibfield  {author} {\bibinfo {author} {\bibfnamefont {J.~M.}\ \bibnamefont
  {Deutsch}},\ }\href@noop {} {\bibfield  {journal} {\bibinfo  {journal}
  {Physical Review A}\ }\textbf {\bibinfo {volume} {43}},\ \bibinfo {pages}
  {2046} (\bibinfo {year} {1991})}\BibitemShut {NoStop}%
\bibitem [{\citenamefont {Srednicki}(1994)}]{srednicki1994chaos}%
  \BibitemOpen
  \bibfield  {author} {\bibinfo {author} {\bibfnamefont {M.}~\bibnamefont
  {Srednicki}},\ }\href@noop {} {\bibfield  {journal} {\bibinfo  {journal}
  {Physical Review E}\ }\textbf {\bibinfo {volume} {50}},\ \bibinfo {pages}
  {888} (\bibinfo {year} {1994})}\BibitemShut {NoStop}%
\bibitem [{\citenamefont {Tasaki}(1998)}]{tasaki1998quantum}%
  \BibitemOpen
  \bibfield  {author} {\bibinfo {author} {\bibfnamefont {H.}~\bibnamefont
  {Tasaki}},\ }\href@noop {} {\bibfield  {journal} {\bibinfo  {journal}
  {Physical review letters}\ }\textbf {\bibinfo {volume} {80}},\ \bibinfo
  {pages} {1373} (\bibinfo {year} {1998})}\BibitemShut {NoStop}%
\bibitem [{\citenamefont {Rigol}\ \emph {et~al.}(2008)\citenamefont {Rigol},
  \citenamefont {Dunjko},\ and\ \citenamefont
  {Olshanii}}]{rigol2008thermalization}%
  \BibitemOpen
  \bibfield  {author} {\bibinfo {author} {\bibfnamefont {M.}~\bibnamefont
  {Rigol}}, \bibinfo {author} {\bibfnamefont {V.}~\bibnamefont {Dunjko}}, \
  and\ \bibinfo {author} {\bibfnamefont {M.}~\bibnamefont {Olshanii}},\
  }\href@noop {} {\bibfield  {journal} {\bibinfo  {journal} {Nature}\ }\textbf
  {\bibinfo {volume} {452}},\ \bibinfo {pages} {854} (\bibinfo {year}
  {2008})}\BibitemShut {NoStop}%
\bibitem [{\citenamefont {Hayden}\ and\ \citenamefont
  {Preskill}(2007)}]{Hayden07}%
  \BibitemOpen
  \bibfield  {author} {\bibinfo {author} {\bibfnamefont {P.}~\bibnamefont
  {Hayden}}\ and\ \bibinfo {author} {\bibfnamefont {J.}~\bibnamefont
  {Preskill}},\ }\href@noop {} {\bibfield  {journal} {\bibinfo  {journal}
  {JHEP}\ }\textbf {\bibinfo {volume} {09}},\ \bibinfo {pages} {120} (\bibinfo
  {year} {2007})}\BibitemShut {NoStop}%
\bibitem [{\citenamefont {Sekino}\ and\ \citenamefont
  {Susskind}(2008)}]{Sekino08}%
  \BibitemOpen
  \bibfield  {author} {\bibinfo {author} {\bibfnamefont {Y.}~\bibnamefont
  {Sekino}}\ and\ \bibinfo {author} {\bibfnamefont {L.}~\bibnamefont
  {Susskind}},\ }\href@noop {} {\bibfield  {journal} {\bibinfo  {journal}
  {JHEP}\ }\textbf {\bibinfo {volume} {10}},\ \bibinfo {pages} {065} (\bibinfo
  {year} {2008})}\BibitemShut {NoStop}%
\bibitem [{\citenamefont {Lashkari}\ \emph {et~al.}(2013)\citenamefont
  {Lashkari}, \citenamefont {Stanford}, \citenamefont {Hastings}, \citenamefont
  {Osborne},\ and\ \citenamefont {Hayden}}]{Lashkari13}%
  \BibitemOpen
  \bibfield  {author} {\bibinfo {author} {\bibfnamefont {N.}~\bibnamefont
  {Lashkari}}, \bibinfo {author} {\bibfnamefont {D.}~\bibnamefont {Stanford}},
  \bibinfo {author} {\bibfnamefont {M.}~\bibnamefont {Hastings}}, \bibinfo
  {author} {\bibfnamefont {T.}~\bibnamefont {Osborne}}, \ and\ \bibinfo
  {author} {\bibfnamefont {P.}~\bibnamefont {Hayden}},\ }\href {\doibase
  10.1007/JHEP04(2013)022} {\bibfield  {journal} {\bibinfo  {journal} {JHEP}\
  }\textbf {\bibinfo {volume} {04}},\ \bibinfo {pages} {22} (\bibinfo {year}
  {2013})}\BibitemShut {NoStop}%
\bibitem [{\citenamefont {Kitaev}(2014)}]{Kitaev:2014t1}%
  \BibitemOpen
  \bibfield  {author} {\bibinfo {author} {\bibfnamefont {A.}~\bibnamefont
  {Kitaev}},\ }\href@noop {} {\enquote {\bibinfo {title} {Hidden correlations
  in the hawking radiation and thermal noise},}\ } (\bibinfo {year} {2014}),\
  \bibinfo {note} {talk given at the Fundamental Physics Prize Symposium, Nov.
  10, 2014}\BibitemShut {NoStop}%
\bibitem [{\citenamefont {Maldacena}\ and\ \citenamefont
  {Stanford}(2016)}]{Maldacena:2016ac}%
  \BibitemOpen
  \bibfield  {author} {\bibinfo {author} {\bibfnamefont {J.}~\bibnamefont
  {Maldacena}}\ and\ \bibinfo {author} {\bibfnamefont {D.}~\bibnamefont
  {Stanford}},\ }\href@noop {} {\bibfield  {journal} {\bibinfo  {journal}
  {Phys. Rev. D}\ }\textbf {\bibinfo {volume} {94}},\ \bibinfo {pages} {106002}
  (\bibinfo {year} {2016})}\BibitemShut {NoStop}%
\bibitem [{\citenamefont {Shenker}\ and\ \citenamefont
  {Stanford}(2014{\natexlab{a}})}]{Shenker:2013pqa}%
  \BibitemOpen
  \bibfield  {author} {\bibinfo {author} {\bibfnamefont {S.~H.}\ \bibnamefont
  {Shenker}}\ and\ \bibinfo {author} {\bibfnamefont {D.}~\bibnamefont
  {Stanford}},\ }\href {\doibase 10.1007/JHEP03(2014)067} {\bibfield  {journal}
  {\bibinfo  {journal} {JHEP}\ }\textbf {\bibinfo {volume} {03}},\ \bibinfo
  {pages} {067} (\bibinfo {year} {2014}{\natexlab{a}})}\BibitemShut {NoStop}%
\bibitem [{\citenamefont {Maldacena}\ \emph {et~al.}(2017)\citenamefont
  {Maldacena}, \citenamefont {Stanford},\ and\ \citenamefont
  {Yang}}]{Traversable2017}%
  \BibitemOpen
  \bibfield  {author} {\bibinfo {author} {\bibfnamefont {J.}~\bibnamefont
  {Maldacena}}, \bibinfo {author} {\bibfnamefont {D.}~\bibnamefont {Stanford}},
  \ and\ \bibinfo {author} {\bibfnamefont {Z.}~\bibnamefont {Yang}},\ }\href
  {\doibase 10.1002/prop.201700034} {\bibfield  {journal} {\bibinfo  {journal}
  {Fortsch. Phys.}\ }\textbf {\bibinfo {volume} {65}},\ \bibinfo {pages}
  {1700034} (\bibinfo {year} {2017})}\BibitemShut {NoStop}%
\bibitem [{\citenamefont {Roberts}\ \emph {et~al.}(2015)\citenamefont
  {Roberts}, \citenamefont {Stanford},\ and\ \citenamefont
  {Susskind}}]{Roberts:2014isa}%
  \BibitemOpen
  \bibfield  {author} {\bibinfo {author} {\bibfnamefont {D.~A.}\ \bibnamefont
  {Roberts}}, \bibinfo {author} {\bibfnamefont {D.}~\bibnamefont {Stanford}}, \
  and\ \bibinfo {author} {\bibfnamefont {L.}~\bibnamefont {Susskind}},\ }\href
  {\doibase 10.1007/JHEP03(2015)051} {\bibfield  {journal} {\bibinfo  {journal}
  {JHEP}\ }\textbf {\bibinfo {volume} {03}},\ \bibinfo {pages} {051} (\bibinfo
  {year} {2015})}\BibitemShut {NoStop}%
\bibitem [{\citenamefont {Shenker}\ and\ \citenamefont
  {Stanford}(2014{\natexlab{b}})}]{Shenker:2013yza}%
  \BibitemOpen
  \bibfield  {author} {\bibinfo {author} {\bibfnamefont {S.~H.}\ \bibnamefont
  {Shenker}}\ and\ \bibinfo {author} {\bibfnamefont {D.}~\bibnamefont
  {Stanford}},\ }\href {\doibase 10.1007/JHEP12(2014)046} {\bibfield  {journal}
  {\bibinfo  {journal} {JHEP}\ }\textbf {\bibinfo {volume} {12}},\ \bibinfo
  {pages} {046} (\bibinfo {year} {2014}{\natexlab{b}})}\BibitemShut {NoStop}%
\bibitem [{\citenamefont {Roberts}\ and\ \citenamefont
  {Stanford}(2015)}]{Roberts:2014ifa}%
  \BibitemOpen
  \bibfield  {author} {\bibinfo {author} {\bibfnamefont {D.~A.}\ \bibnamefont
  {Roberts}}\ and\ \bibinfo {author} {\bibfnamefont {D.}~\bibnamefont
  {Stanford}},\ }\href {\doibase 10.1103/PhysRevLett.115.131603} {\bibfield
  {journal} {\bibinfo  {journal} {Phys. Rev. Lett.}\ }\textbf {\bibinfo
  {volume} {115}},\ \bibinfo {pages} {131603} (\bibinfo {year}
  {2015})}\BibitemShut {NoStop}%
\bibitem [{\citenamefont {Hosur}\ \emph {et~al.}(2016)\citenamefont {Hosur},
  \citenamefont {Qi}, \citenamefont {Roberts},\ and\ \citenamefont
  {Yoshida}}]{Hosur:2015ylk}%
  \BibitemOpen
  \bibfield  {author} {\bibinfo {author} {\bibfnamefont {P.}~\bibnamefont
  {Hosur}}, \bibinfo {author} {\bibfnamefont {X.-L.}\ \bibnamefont {Qi}},
  \bibinfo {author} {\bibfnamefont {D.~A.}\ \bibnamefont {Roberts}}, \ and\
  \bibinfo {author} {\bibfnamefont {B.}~\bibnamefont {Yoshida}},\ }\href
  {\doibase 10.1007/JHEP02(2016)004} {\bibfield  {journal} {\bibinfo  {journal}
  {JHEP}\ }\textbf {\bibinfo {volume} {02}},\ \bibinfo {pages} {004} (\bibinfo
  {year} {2016})}\BibitemShut {NoStop}%
\bibitem [{\citenamefont {Roberts}\ and\ \citenamefont
  {Yoshida}(2017)}]{Roberts:2017aa}%
  \BibitemOpen
  \bibfield  {author} {\bibinfo {author} {\bibfnamefont {D.~A.}\ \bibnamefont
  {Roberts}}\ and\ \bibinfo {author} {\bibfnamefont {B.}~\bibnamefont
  {Yoshida}},\ }\href {\doibase 10.1007/JHEP04(2017)121} {\bibfield  {journal}
  {\bibinfo  {journal} {JHEP}\ }\textbf {\bibinfo {volume} {04}},\ \bibinfo
  {pages} {121} (\bibinfo {year} {2017})}\BibitemShut {NoStop}%
\bibitem [{\citenamefont {Blake}(2016{\natexlab{a}})}]{Blake:2016wvh}%
  \BibitemOpen
  \bibfield  {author} {\bibinfo {author} {\bibfnamefont {M.}~\bibnamefont
  {Blake}},\ }\href {\doibase 10.1103/PhysRevLett.117.091601} {\bibfield
  {journal} {\bibinfo  {journal} {Phys. Rev. Lett.}\ }\textbf {\bibinfo
  {volume} {117}},\ \bibinfo {pages} {091601} (\bibinfo {year}
  {2016}{\natexlab{a}})}\BibitemShut {NoStop}%
\bibitem [{\citenamefont {Blake}(2016{\natexlab{b}})}]{Blake:2016aa}%
  \BibitemOpen
  \bibfield  {author} {\bibinfo {author} {\bibfnamefont {M.}~\bibnamefont
  {Blake}},\ }\href {https://link.aps.org/doi/10.1103/PhysRevD.94.086014}
  {\bibfield  {journal} {\bibinfo  {journal} {Phys. Rev. D}\ }\textbf {\bibinfo
  {volume} {94}},\ \bibinfo {pages} {086014} (\bibinfo {year}
  {2016}{\natexlab{b}})}\BibitemShut {NoStop}%
\bibitem [{\citenamefont {Kitaev}(2015)}]{Kitaev:2014t2}%
  \BibitemOpen
  \bibfield  {author} {\bibinfo {author} {\bibfnamefont {A.}~\bibnamefont
  {Kitaev}},\ }\href@noop {} {\enquote {\bibinfo {title} {A simple model of
  quantum holography},}\ } (\bibinfo {year} {2015}),\ \bibinfo {note} {talks at
  KITP, April 7, 2015 and May 27, 2015}\BibitemShut {NoStop}%
\bibitem [{\citenamefont {Nahum}\ \emph {et~al.}(2017)\citenamefont {Nahum},
  \citenamefont {Ruhman}, \citenamefont {Vijay},\ and\ \citenamefont
  {Haah}}]{Nahum:2017aa}%
  \BibitemOpen
  \bibfield  {author} {\bibinfo {author} {\bibfnamefont {A.}~\bibnamefont
  {Nahum}}, \bibinfo {author} {\bibfnamefont {J.}~\bibnamefont {Ruhman}},
  \bibinfo {author} {\bibfnamefont {S.}~\bibnamefont {Vijay}}, \ and\ \bibinfo
  {author} {\bibfnamefont {J.}~\bibnamefont {Haah}},\ }\href@noop {} {\bibfield
   {journal} {\bibinfo  {journal} {Phys. Rev. X}\ }\textbf {\bibinfo {volume}
  {7}},\ \bibinfo {pages} {031016} (\bibinfo {year} {2017})}\BibitemShut
  {NoStop}%
\bibitem [{\citenamefont {von Keyserlingk}\ \emph {et~al.}(2017)\citenamefont
  {von Keyserlingk}, \citenamefont {Rakovszky}, \citenamefont {Pollmann},\ and\
  \citenamefont {Sondhi}}]{Keyserlingk:2017}%
  \BibitemOpen
  \bibfield  {author} {\bibinfo {author} {\bibfnamefont {C.}~\bibnamefont {von
  Keyserlingk}}, \bibinfo {author} {\bibfnamefont {T.}~\bibnamefont
  {Rakovszky}}, \bibinfo {author} {\bibfnamefont {F.}~\bibnamefont {Pollmann}},
  \ and\ \bibinfo {author} {\bibfnamefont {S.}~\bibnamefont {Sondhi}},\
  }\href@noop {} {\  (\bibinfo {year} {2017})},\ \Eprint
  {http://arxiv.org/abs/arXiv:1705.08910} {arXiv:1705.08910} \BibitemShut
  {NoStop}%
\bibitem [{\citenamefont {Cotler}\ \emph {et~al.}(2017)\citenamefont {Cotler},
  \citenamefont {Hunter-Jones}, \citenamefont {Liu},\ and\ \citenamefont
  {Yoshida}}]{Cotler:2017aa}%
  \BibitemOpen
  \bibfield  {author} {\bibinfo {author} {\bibfnamefont {J.}~\bibnamefont
  {Cotler}}, \bibinfo {author} {\bibfnamefont {N.}~\bibnamefont
  {Hunter-Jones}}, \bibinfo {author} {\bibfnamefont {J.}~\bibnamefont {Liu}}, \
  and\ \bibinfo {author} {\bibfnamefont {B.}~\bibnamefont {Yoshida}},\ }\href
  {\doibase 10.1007/JHEP11(2017)048} {\bibfield  {journal} {\bibinfo  {journal}
  {JHEP}\ }\textbf {\bibinfo {volume} {11}},\ \bibinfo {pages} {48} (\bibinfo
  {year} {2017})}\BibitemShut {NoStop}%
\bibitem [{\citenamefont {Khemani}\ \emph {et~al.}()\citenamefont {Khemani},
  \citenamefont {Vishwanath},\ and\ \citenamefont {Huse}}]{Khemani:2017aa}%
  \BibitemOpen
  \bibfield  {author} {\bibinfo {author} {\bibfnamefont {V.}~\bibnamefont
  {Khemani}}, \bibinfo {author} {\bibfnamefont {A.}~\bibnamefont {Vishwanath}},
  \ and\ \bibinfo {author} {\bibfnamefont {D.~A.}\ \bibnamefont {Huse}},\
  }\href {https://arxiv.org/abs/1710.09835} {\ }\Eprint
  {http://arxiv.org/abs/arXiv:1710.09835} {arXiv:1710.09835} \BibitemShut
  {NoStop}%
\bibitem [{\citenamefont {Davison}\ \emph {et~al.}(2017)\citenamefont
  {Davison}, \citenamefont {Fu}, \citenamefont {Georges}, \citenamefont {Gu},
  \citenamefont {Jensen},\ and\ \citenamefont {Sachdev}}]{Davison:2017aa}%
  \BibitemOpen
  \bibfield  {author} {\bibinfo {author} {\bibfnamefont {R.~A.}\ \bibnamefont
  {Davison}}, \bibinfo {author} {\bibfnamefont {W.}~\bibnamefont {Fu}},
  \bibinfo {author} {\bibfnamefont {A.}~\bibnamefont {Georges}}, \bibinfo
  {author} {\bibfnamefont {Y.}~\bibnamefont {Gu}}, \bibinfo {author}
  {\bibfnamefont {K.}~\bibnamefont {Jensen}}, \ and\ \bibinfo {author}
  {\bibfnamefont {S.}~\bibnamefont {Sachdev}},\ }\href
  {https://link.aps.org/doi/10.1103/PhysRevB.95.155131} {\bibfield  {journal}
  {\bibinfo  {journal} {Phys. Rev. B}\ }\textbf {\bibinfo {volume} {95}},\
  \bibinfo {pages} {155131} (\bibinfo {year} {2017})}\BibitemShut {NoStop}%
\bibitem [{\citenamefont {Gu}\ \emph {et~al.}(2017)\citenamefont {Gu},
  \citenamefont {Qi},\ and\ \citenamefont {Stanford}}]{Gu:2017aa}%
  \BibitemOpen
  \bibfield  {author} {\bibinfo {author} {\bibfnamefont {Y.}~\bibnamefont
  {Gu}}, \bibinfo {author} {\bibfnamefont {X.-L.}\ \bibnamefont {Qi}}, \ and\
  \bibinfo {author} {\bibfnamefont {D.}~\bibnamefont {Stanford}},\ }\href
  {\doibase 10.1007/JHEP05(2017)125} {\bibfield  {journal} {\bibinfo  {journal}
  {JHEP}\ }\textbf {\bibinfo {volume} {05}},\ \bibinfo {pages} {125} (\bibinfo
  {year} {2017})}\BibitemShut {NoStop}%
\bibitem [{\citenamefont {Gao}\ \emph {et~al.}(2016)\citenamefont {Gao},
  \citenamefont {Jafferis},\ and\ \citenamefont {Wall}}]{Gao:2016aa}%
  \BibitemOpen
  \bibfield  {author} {\bibinfo {author} {\bibfnamefont {P.}~\bibnamefont
  {Gao}}, \bibinfo {author} {\bibfnamefont {D.~L.}\ \bibnamefont {Jafferis}}, \
  and\ \bibinfo {author} {\bibfnamefont {A.}~\bibnamefont {Wall}},\ }\href@noop
  {} {\  (\bibinfo {year} {2016})},\ \Eprint
  {http://arxiv.org/abs/arXiv:1608.05687} {arXiv:1608.05687} \BibitemShut
  {NoStop}%
\bibitem [{\citenamefont {Page}(1993)}]{Page93}%
  \BibitemOpen
  \bibfield  {author} {\bibinfo {author} {\bibfnamefont {D.~N.}\ \bibnamefont
  {Page}},\ }\href@noop {} {\bibfield  {journal} {\bibinfo  {journal} {Phys.
  Rev. Lett.}\ }\textbf {\bibinfo {volume} {71}},\ \bibinfo {pages} {1291}
  (\bibinfo {year} {1993})}\BibitemShut {NoStop}%
\bibitem [{\citenamefont {Yoshida}\ and\ \citenamefont
  {Kitaev}(2017)}]{Yoshida:2017aa}%
  \BibitemOpen
  \bibfield  {author} {\bibinfo {author} {\bibfnamefont {B.}~\bibnamefont
  {Yoshida}}\ and\ \bibinfo {author} {\bibfnamefont {A.}~\bibnamefont
  {Kitaev}},\ }\href {https://arxiv.org/abs/1710.03363} {\  (\bibinfo {year}
  {2017})},\ \Eprint {http://arxiv.org/abs/arXiv:1710.03363} {arXiv:1710.03363}
  \BibitemShut {NoStop}%
\bibitem [{\citenamefont {Banerjee}\ and\ \citenamefont
  {Altman}(2017)}]{banerjee2017solvable}%
  \BibitemOpen
  \bibfield  {author} {\bibinfo {author} {\bibfnamefont {S.}~\bibnamefont
  {Banerjee}}\ and\ \bibinfo {author} {\bibfnamefont {E.}~\bibnamefont
  {Altman}},\ }\href@noop {} {\bibfield  {journal} {\bibinfo  {journal}
  {Physical Review B}\ }\textbf {\bibinfo {volume} {95}},\ \bibinfo {pages}
  {134302} (\bibinfo {year} {2017})}\BibitemShut {NoStop}%
\bibitem [{\citenamefont {Patel}\ and\ \citenamefont
  {Sachdev}(2017)}]{patel2017quantum}%
  \BibitemOpen
  \bibfield  {author} {\bibinfo {author} {\bibfnamefont {A.~A.}\ \bibnamefont
  {Patel}}\ and\ \bibinfo {author} {\bibfnamefont {S.}~\bibnamefont
  {Sachdev}},\ }\href@noop {} {\bibfield  {journal} {\bibinfo  {journal}
  {Proceedings of the National Academy of Sciences}\ }\textbf {\bibinfo
  {volume} {114}},\ \bibinfo {pages} {1844} (\bibinfo {year}
  {2017})}\BibitemShut {NoStop}%
\bibitem [{\citenamefont {Larkin}\ and\ \citenamefont
  {Ovchinnikov}(1969)}]{larkin1969quasiclassical}%
  \BibitemOpen
  \bibfield  {author} {\bibinfo {author} {\bibfnamefont {A.}~\bibnamefont
  {Larkin}}\ and\ \bibinfo {author} {\bibfnamefont {Y.~N.}\ \bibnamefont
  {Ovchinnikov}},\ }\href@noop {} {\bibfield  {journal} {\bibinfo  {journal}
  {Sov Phys JETP}\ }\textbf {\bibinfo {volume} {28}},\ \bibinfo {pages} {1200}
  (\bibinfo {year} {1969})}\BibitemShut {NoStop}%
\bibitem [{\citenamefont {Swingle}\ \emph {et~al.}(2016)\citenamefont
  {Swingle}, \citenamefont {Bentsen}, \citenamefont {Schleier-Smith},\ and\
  \citenamefont {Hayden}}]{swingle2016measuring}%
  \BibitemOpen
  \bibfield  {author} {\bibinfo {author} {\bibfnamefont {B.}~\bibnamefont
  {Swingle}}, \bibinfo {author} {\bibfnamefont {G.}~\bibnamefont {Bentsen}},
  \bibinfo {author} {\bibfnamefont {M.}~\bibnamefont {Schleier-Smith}}, \ and\
  \bibinfo {author} {\bibfnamefont {P.}~\bibnamefont {Hayden}},\ }\href@noop {}
  {\bibfield  {journal} {\bibinfo  {journal} {Physical Review A}\ }\textbf
  {\bibinfo {volume} {94}},\ \bibinfo {pages} {040302} (\bibinfo {year}
  {2016})}\BibitemShut {NoStop}%
\bibitem [{\citenamefont {Yao}\ \emph {et~al.}(2016)\citenamefont {Yao},
  \citenamefont {Grusdt}, \citenamefont {Swingle}, \citenamefont {Lukin},
  \citenamefont {Stamper-Kurn}, \citenamefont {Moore},\ and\ \citenamefont
  {Demler}}]{yao2016interferometric}%
  \BibitemOpen
  \bibfield  {author} {\bibinfo {author} {\bibfnamefont {N.~Y.}\ \bibnamefont
  {Yao}}, \bibinfo {author} {\bibfnamefont {F.}~\bibnamefont {Grusdt}},
  \bibinfo {author} {\bibfnamefont {B.}~\bibnamefont {Swingle}}, \bibinfo
  {author} {\bibfnamefont {M.~D.}\ \bibnamefont {Lukin}}, \bibinfo {author}
  {\bibfnamefont {D.~M.}\ \bibnamefont {Stamper-Kurn}}, \bibinfo {author}
  {\bibfnamefont {J.~E.}\ \bibnamefont {Moore}}, \ and\ \bibinfo {author}
  {\bibfnamefont {E.~A.}\ \bibnamefont {Demler}},\ }\href@noop {} {\bibfield
  {journal} {\bibinfo  {journal} {arXiv preprint arXiv:1607.01801}\ } (\bibinfo
  {year} {2016})}\BibitemShut {NoStop}%
\bibitem [{\citenamefont {Zhu}\ \emph {et~al.}(2016)\citenamefont {Zhu},
  \citenamefont {Hafezi},\ and\ \citenamefont {Grover}}]{zhu2016measurement}%
  \BibitemOpen
  \bibfield  {author} {\bibinfo {author} {\bibfnamefont {G.}~\bibnamefont
  {Zhu}}, \bibinfo {author} {\bibfnamefont {M.}~\bibnamefont {Hafezi}}, \ and\
  \bibinfo {author} {\bibfnamefont {T.}~\bibnamefont {Grover}},\ }\href@noop {}
  {\bibfield  {journal} {\bibinfo  {journal} {Physical Review A}\ }\textbf
  {\bibinfo {volume} {94}},\ \bibinfo {pages} {062329} (\bibinfo {year}
  {2016})}\BibitemShut {NoStop}%
\bibitem [{\citenamefont {G{\"a}rttner}\ \emph {et~al.}(2017)\citenamefont
  {G{\"a}rttner}, \citenamefont {Bohnet}, \citenamefont {Safavi-Naini},
  \citenamefont {Wall}, \citenamefont {Bollinger},\ and\ \citenamefont
  {Rey}}]{garttner2017measuring}%
  \BibitemOpen
  \bibfield  {author} {\bibinfo {author} {\bibfnamefont {M.}~\bibnamefont
  {G{\"a}rttner}}, \bibinfo {author} {\bibfnamefont {J.~G.}\ \bibnamefont
  {Bohnet}}, \bibinfo {author} {\bibfnamefont {A.}~\bibnamefont
  {Safavi-Naini}}, \bibinfo {author} {\bibfnamefont {M.~L.}\ \bibnamefont
  {Wall}}, \bibinfo {author} {\bibfnamefont {J.~J.}\ \bibnamefont {Bollinger}},
  \ and\ \bibinfo {author} {\bibfnamefont {A.~M.}\ \bibnamefont {Rey}},\
  }\href@noop {} {\bibfield  {journal} {\bibinfo  {journal} {Nature Physics}\ }
  (\bibinfo {year} {2017})}\BibitemShut {NoStop}%
\bibitem [{\citenamefont {Li}\ \emph {et~al.}(2017)\citenamefont {Li},
  \citenamefont {Fan}, \citenamefont {Wang}, \citenamefont {Ye}, \citenamefont
  {Zeng}, \citenamefont {Zhai}, \citenamefont {Peng},\ and\ \citenamefont
  {Du}}]{li2017measuring}%
  \BibitemOpen
  \bibfield  {author} {\bibinfo {author} {\bibfnamefont {J.}~\bibnamefont
  {Li}}, \bibinfo {author} {\bibfnamefont {R.}~\bibnamefont {Fan}}, \bibinfo
  {author} {\bibfnamefont {H.}~\bibnamefont {Wang}}, \bibinfo {author}
  {\bibfnamefont {B.}~\bibnamefont {Ye}}, \bibinfo {author} {\bibfnamefont
  {B.}~\bibnamefont {Zeng}}, \bibinfo {author} {\bibfnamefont {H.}~\bibnamefont
  {Zhai}}, \bibinfo {author} {\bibfnamefont {X.}~\bibnamefont {Peng}}, \ and\
  \bibinfo {author} {\bibfnamefont {J.}~\bibnamefont {Du}},\ }\href@noop {}
  {\bibfield  {journal} {\bibinfo  {journal} {Physical Review X}\ }\textbf
  {\bibinfo {volume} {7}},\ \bibinfo {pages} {031011} (\bibinfo {year}
  {2017})}\BibitemShut {NoStop}%
\bibitem [{\citenamefont {Hradil}(1997)}]{hradil1997quantum}%
  \BibitemOpen
  \bibfield  {author} {\bibinfo {author} {\bibfnamefont {Z.}~\bibnamefont
  {Hradil}},\ }\href@noop {} {\bibfield  {journal} {\bibinfo  {journal}
  {Physical Review A}\ }\textbf {\bibinfo {volume} {55}},\ \bibinfo {pages}
  {R1561} (\bibinfo {year} {1997})}\BibitemShut {NoStop}%
\bibitem [{\citenamefont {Dodonov}\ and\ \citenamefont
  {Man'ko}(1997)}]{dodonov1997positive}%
  \BibitemOpen
  \bibfield  {author} {\bibinfo {author} {\bibfnamefont {V.}~\bibnamefont
  {Dodonov}}\ and\ \bibinfo {author} {\bibfnamefont {V.}~\bibnamefont
  {Man'ko}},\ }\href@noop {} {\bibfield  {journal} {\bibinfo  {journal}
  {Physics Letters A}\ }\textbf {\bibinfo {volume} {229}},\ \bibinfo {pages}
  {335} (\bibinfo {year} {1997})}\BibitemShut {NoStop}%
\bibitem [{\citenamefont {H{\"a}ffner}\ \emph {et~al.}(2005)\citenamefont
  {H{\"a}ffner}, \citenamefont {H{\"a}nsel}, \citenamefont {Roos},
  \citenamefont {Benhelm}, \citenamefont {Chwalla}, \citenamefont {K{\"o}rber},
  \citenamefont {Rapol}, \citenamefont {Riebe}, \citenamefont {Schmidt},
  \citenamefont {Becher} \emph {et~al.}}]{haffner2005scalable}%
  \BibitemOpen
  \bibfield  {author} {\bibinfo {author} {\bibfnamefont {H.}~\bibnamefont
  {H{\"a}ffner}}, \bibinfo {author} {\bibfnamefont {W.}~\bibnamefont
  {H{\"a}nsel}}, \bibinfo {author} {\bibfnamefont {C.}~\bibnamefont {Roos}},
  \bibinfo {author} {\bibfnamefont {J.}~\bibnamefont {Benhelm}}, \bibinfo
  {author} {\bibfnamefont {M.}~\bibnamefont {Chwalla}}, \bibinfo {author}
  {\bibfnamefont {T.}~\bibnamefont {K{\"o}rber}}, \bibinfo {author}
  {\bibfnamefont {U.}~\bibnamefont {Rapol}}, \bibinfo {author} {\bibfnamefont
  {M.}~\bibnamefont {Riebe}}, \bibinfo {author} {\bibfnamefont
  {P.}~\bibnamefont {Schmidt}}, \bibinfo {author} {\bibfnamefont
  {C.}~\bibnamefont {Becher}},  \emph {et~al.},\ }\href@noop {} {\bibfield
  {journal} {\bibinfo  {journal} {Nature}\ }\textbf {\bibinfo {volume} {438}},\
  \bibinfo {pages} {643} (\bibinfo {year} {2005})}\BibitemShut {NoStop}%
\bibitem [{\citenamefont {Gross}\ \emph {et~al.}(2010)\citenamefont {Gross},
  \citenamefont {Liu}, \citenamefont {Flammia}, \citenamefont {Becker},\ and\
  \citenamefont {Eisert}}]{gross2010quantum}%
  \BibitemOpen
  \bibfield  {author} {\bibinfo {author} {\bibfnamefont {D.}~\bibnamefont
  {Gross}}, \bibinfo {author} {\bibfnamefont {Y.-K.}\ \bibnamefont {Liu}},
  \bibinfo {author} {\bibfnamefont {S.~T.}\ \bibnamefont {Flammia}}, \bibinfo
  {author} {\bibfnamefont {S.}~\bibnamefont {Becker}}, \ and\ \bibinfo {author}
  {\bibfnamefont {J.}~\bibnamefont {Eisert}},\ }\href@noop {} {\bibfield
  {journal} {\bibinfo  {journal} {Physical review letters}\ }\textbf {\bibinfo
  {volume} {105}},\ \bibinfo {pages} {150401} (\bibinfo {year}
  {2010})}\BibitemShut {NoStop}%
\bibitem [{\citenamefont {Grover}(1996)}]{grover1996fast}%
  \BibitemOpen
  \bibfield  {author} {\bibinfo {author} {\bibfnamefont {L.~K.}\ \bibnamefont
  {Grover}},\ }in\ \href@noop {} {\emph {\bibinfo {booktitle} {Proceedings of
  the twenty-eighth annual ACM symposium on Theory of computing}}}\ (\bibinfo
  {organization} {ACM},\ \bibinfo {year} {1996})\ pp.\ \bibinfo {pages}
  {212--219}\BibitemShut {NoStop}%
\bibitem [{\citenamefont {Huang}\ \emph {et~al.}()\citenamefont {Huang},
  \citenamefont {Brandao},\ and\ \citenamefont {Zhang}}]{Huang:2017aa}%
  \BibitemOpen
  \bibfield  {author} {\bibinfo {author} {\bibfnamefont {Y.}~\bibnamefont
  {Huang}}, \bibinfo {author} {\bibfnamefont {F.~G.}\ \bibnamefont {Brandao}},
  \ and\ \bibinfo {author} {\bibfnamefont {Y.-L.}\ \bibnamefont {Zhang}},\
  }\href@noop {} {}\Eprint {http://arxiv.org/abs/arXiv:1705.07597}
  {arXiv:1705.07597} \BibitemShut {NoStop}%
\bibitem [{Note1()}]{Note1}%
  \BibitemOpen
  \bibinfo {note} {Eq.~\protect \textup {\hbox {\mathsurround \z@ \protect
  \normalfont (\ignorespaces \ref {eq:1-design}\unskip \@@italiccorr )}} holds
  since the Pauli operators form a unitary $1$-design \cite
  {renes2004symmetric,dankert2009exact}. Note that there are $d_{A}^2,d_{D}^2$
  Pauli operators (including the identity operator) on regions $A,D$,
  respectively.}\BibitemShut {Stop}%
\bibitem [{Note2()}]{Note2}%
  \BibitemOpen
  \bibinfo {note} {We note that there are unitary operators which satisfy
  Eq.~\protect \textup {\hbox {\mathsurround \z@ \protect \normalfont
  (\ignorespaces \ref {eq:ave-def}\unskip \@@italiccorr )}}, but not
  Eq.~\protect \textup {\hbox {\mathsurround \z@ \protect \normalfont
  (\ignorespaces \ref {eq:def}\unskip \@@italiccorr )}}. For example, a random
  Clifford operator is scrambling for Eq.~\protect \textup {\hbox
  {\mathsurround \z@ \protect \normalfont (\ignorespaces \ref
  {eq:ave-def}\unskip \@@italiccorr )}}, since the Clifford operators form a
  unitary $2$-design. However, OTOCs for a Clifford unitary are always $\pm 1$
  if $O_{X}=O_{Z}$ and $O_{Y}=O_{W}$ are Pauli operators, and thus do not
  satisfy Eq.~\protect \textup {\hbox {\mathsurround \z@ \protect \normalfont
  (\ignorespaces \ref {eq:def}\unskip \@@italiccorr )}}. In this sense, a
  random unitary from a $2$-design is not enough to achieve full scrambling.
  Rather, to achieve full scrambling, it suffices to pick a random operator $U$
  from a unitary $4$-design.}\BibitemShut {Stop}%
\bibitem [{sup()}]{suppinfo}%
  \BibitemOpen
  \href@noop {} {\bibinfo  {journal} {See supplementary information}\
  }\BibitemShut {NoStop}%
\bibitem [{\citenamefont {Hayden}\ \emph {et~al.}(2008)\citenamefont {Hayden},
  \citenamefont {Horodecki}, \citenamefont {Winter},\ and\ \citenamefont
  {Yard}}]{Hayden:2008aa}%
  \BibitemOpen
\bibfield  {journal} {  }\bibfield  {author} {\bibinfo {author} {\bibfnamefont
  {P.}~\bibnamefont {Hayden}}, \bibinfo {author} {\bibfnamefont
  {M.}~\bibnamefont {Horodecki}}, \bibinfo {author} {\bibfnamefont
  {A.}~\bibnamefont {Winter}}, \ and\ \bibinfo {author} {\bibfnamefont
  {J.}~\bibnamefont {Yard}},\ }\href {\doibase 10.1142/S1230161208000043}
  {\bibfield  {journal} {\bibinfo  {journal} {Open Syst. Inf. Dyn.}\ }\textbf
  {\bibinfo {volume} {15}},\ \bibinfo {pages} {7} (\bibinfo {year}
  {2008})}\BibitemShut {NoStop}%
\bibitem [{Note3()}]{Note3}%
  \BibitemOpen
  \bibinfo {note} {It is possible to generalize our results to finite
  temperature (factorable ensembles) using the R\'{e}nyi divergence \cite
  {suppinfo}.}\BibitemShut {Stop}%
\bibitem [{Note4()}]{Note4}%
  \BibitemOpen
  \bibinfo {note} {While the R\'{e}nyi-$2$ mutual information is a measurable
  quantity as the average of OTOCs, the standard mutual information ($\alpha
  =1$) is often more convenient as it satisfies useful monotonicity
  inequalities. For the case of maximally mixed ensembles $\rho =\protect \frac
  {1}{d}I$, one can derive $I(A,B'D)\geq I^{(2)}(A,B'D)$ using the monotonicity
  of R\'{e}nyi entropy. This analysis can be generalized to cases where the
  input and output ensembles factorize \cite {suppinfo}; $\rho _{AB}=\rho
  _{A}\otimes \rho _{B}$ and $\rho _{CD}=\rho _{C}\otimes \rho _{D}$ where the
  R\'{e}nyi-$2$ mutual information is replaced with a certain expression
  involving the R\'{e}nyi-$2$ divergence from which the standard mutual
  information can be lower bounded. See appendix for details.}\BibitemShut
  {Stop}%
\bibitem [{\citenamefont {Nielsen}\ and\ \citenamefont
  {Chuang}(2010)}]{nielsen2010quantum}%
  \BibitemOpen
  \bibfield  {author} {\bibinfo {author} {\bibfnamefont {M.~A.}\ \bibnamefont
  {Nielsen}}\ and\ \bibinfo {author} {\bibfnamefont {I.~L.}\ \bibnamefont
  {Chuang}},\ }\href@noop {} {\emph {\bibinfo {title} {Quantum computation and
  quantum information}}}\ (\bibinfo  {publisher} {Cambridge university press},\
  \bibinfo {year} {2010})\BibitemShut {NoStop}%
\bibitem [{Note5()}]{Note5}%
  \BibitemOpen
  \bibinfo {note} {We note that in the presence of arbitrary forms of noise and
  decoherence, the experimentally ``measured'' value of the OTOC may depend on
  the specific measurement protocol. For example, the OTOC measured via
  interferometric protocols \cite
  {swingle2016measuring,yao2016interferometric,zhu2016measurement,garttner2017measuring,li2017measuring}
  will generically differ from the OTOC measured via our teleportation
  protocol. However, for the important case of a purely depolarizing channel as
  per Eqn.~\protect \textup {\hbox {\mathsurround \z@ \protect \normalfont
  (\ignorespaces \ref {eq:noise}\unskip \@@italiccorr )}}, all such protocols
  will measure the same OTOC given by Eqn.~\protect \textup {\hbox
  {\mathsurround \z@ \protect \normalfont (\ignorespaces \ref
  {eq:OTOCdefn}\unskip \@@italiccorr )}}. Moreover, while the quantitative
  values of noisy OTOCs may differ between protocols, their qualitative decay
  in the presence of decoherence is generic. To this end, a key difference
  between our decoding protocol and previously proposed interferometric
  protocols is the initial preparation of EPR pairs; this preparation is not
  present in the case of interferometric protocols and underlies the reason why
  our teleportation-based method can verify the existence of scrambling
  dynamics while prior methods cannot.}\BibitemShut {Stop}%
\bibitem [{Note6()}]{Note6}%
  \BibitemOpen
  \bibinfo {note} {We do not know the terminology for the expression
  $S_{BD}^{(2)} + S_{D}^{(2)} - S_{B}^{(2)}$, but it is worth noting that
  $S_{BD} + S_{D} - S_{B} \geq 0$ corresponds to the celebrated Araki-Lieb
  inequality.}\BibitemShut {Stop}%
\bibitem [{Note7()}]{Note7}%
  \BibitemOpen
  \bibinfo {note} {To derive this lower bound, we again insert an EPR projector
  on $CC'$ into the diagram for $P_{\psi }F_{\psi }$.}\BibitemShut {Stop}%
\bibitem [{Note8()}]{Note8}%
  \BibitemOpen
  \bibinfo {note} {This integral over $|\psi \protect \rangle $ can be replaced
  with an average over a set of states that form a $2$-design. One example is
  the set of eigenstates of the Pauli operators.}\BibitemShut {Stop}%
\bibitem [{Note9()}]{Note9}%
  \BibitemOpen
  \bibinfo {note} {Note that coherent errors in the initial EPR preparation can
  also be absorbed into the definition of $V$.}\BibitemShut {Stop}%
\bibitem [{Note10()}]{Note10}%
  \BibitemOpen
  \bibinfo {note} {Note that this normalization condition is implied by the
  unitarity of $E$.}\BibitemShut {Stop}%
\bibitem [{Note11()}]{Note11}%
  \BibitemOpen
  \bibinfo {note} {The intuition behind this ambiguity is that for a generic
  quantum channel, one can decompose its action using Kraus operators but this
  decomposition is not unique \cite {nielsen2010quantum}.}\BibitemShut {Stop}%
\bibitem [{Note12()}]{Note12}%
  \BibitemOpen
  \bibinfo {note} {The key point here is that $F_{\protect \text {EPR}}$ always
  provides a lower bound on the \protect \emph {mutual information} between $R$
  and $B'D$, regardless of the nature of experimental errors. Stated
  differently, in the context of the black hole information problem, the fact
  that one can retrieve a quantum state from the Hawking radiation (i.e.~the
  teleportation is successful) implies that the system has scrambled,
  regardless of how one performs the decoding.}\BibitemShut {Stop}%
\bibitem [{Note13()}]{Note13}%
  \BibitemOpen
  \bibinfo {note} {Note that we are using the von Neumann mutual information
  here.}\BibitemShut {Stop}%
\bibitem [{Note14()}]{Note14}%
  \BibitemOpen
  \bibinfo {note} {Recall that R\'{e}nyi-$2$ mutual information is not
  monotonically decreasing in general \cite {suppinfo}.}\BibitemShut {Stop}%
\bibitem [{Note15()}]{Note15}%
  \BibitemOpen
  \bibinfo {note} {Notable examples of fast quantum information scramblers
  include: the SYK model~\cite {Kitaev:2014t2}, $k$-local random spin
  models~\cite {Erdos:2014aa} and random quantum circuits~\cite
  {Dankert09}.}\BibitemShut {Stop}%
\bibitem [{\citenamefont {Zhang}\ \emph {et~al.}(2017)\citenamefont {Zhang},
  \citenamefont {Pagano}, \citenamefont {Hess}, \citenamefont {Kyprianidis},
  \citenamefont {Becker}, \citenamefont {Kaplan}, \citenamefont {Gorshkov},
  \citenamefont {Gong},\ and\ \citenamefont {Monroe}}]{zhang2017observation}%
  \BibitemOpen
  \bibfield  {author} {\bibinfo {author} {\bibfnamefont {J.}~\bibnamefont
  {Zhang}}, \bibinfo {author} {\bibfnamefont {G.}~\bibnamefont {Pagano}},
  \bibinfo {author} {\bibfnamefont {P.~W.}\ \bibnamefont {Hess}}, \bibinfo
  {author} {\bibfnamefont {A.}~\bibnamefont {Kyprianidis}}, \bibinfo {author}
  {\bibfnamefont {P.}~\bibnamefont {Becker}}, \bibinfo {author} {\bibfnamefont
  {H.}~\bibnamefont {Kaplan}}, \bibinfo {author} {\bibfnamefont {A.~V.}\
  \bibnamefont {Gorshkov}}, \bibinfo {author} {\bibfnamefont {Z.-X.}\
  \bibnamefont {Gong}}, \ and\ \bibinfo {author} {\bibfnamefont
  {C.}~\bibnamefont {Monroe}},\ }\href@noop {} {\bibfield  {journal} {\bibinfo
  {journal} {Nature}\ }\textbf {\bibinfo {volume} {551}},\ \bibinfo {pages}
  {601} (\bibinfo {year} {2017})}\BibitemShut {NoStop}%
\bibitem [{\citenamefont {Bernien}\ \emph {et~al.}(2017)\citenamefont
  {Bernien}, \citenamefont {Schwartz}, \citenamefont {Keesling}, \citenamefont
  {Levine}, \citenamefont {Omran}, \citenamefont {Pichler}, \citenamefont
  {Choi}, \citenamefont {Zibrov}, \citenamefont {Endres}, \citenamefont
  {Greiner} \emph {et~al.}}]{bernien2017probing}%
  \BibitemOpen
  \bibfield  {author} {\bibinfo {author} {\bibfnamefont {H.}~\bibnamefont
  {Bernien}}, \bibinfo {author} {\bibfnamefont {S.}~\bibnamefont {Schwartz}},
  \bibinfo {author} {\bibfnamefont {A.}~\bibnamefont {Keesling}}, \bibinfo
  {author} {\bibfnamefont {H.}~\bibnamefont {Levine}}, \bibinfo {author}
  {\bibfnamefont {A.}~\bibnamefont {Omran}}, \bibinfo {author} {\bibfnamefont
  {H.}~\bibnamefont {Pichler}}, \bibinfo {author} {\bibfnamefont
  {S.}~\bibnamefont {Choi}}, \bibinfo {author} {\bibfnamefont {A.~S.}\
  \bibnamefont {Zibrov}}, \bibinfo {author} {\bibfnamefont {M.}~\bibnamefont
  {Endres}}, \bibinfo {author} {\bibfnamefont {M.}~\bibnamefont {Greiner}},
  \emph {et~al.},\ }\href@noop {} {\bibfield  {journal} {\bibinfo  {journal}
  {Nature}\ }\textbf {\bibinfo {volume} {551}},\ \bibinfo {pages} {579}
  (\bibinfo {year} {2017})}\BibitemShut {NoStop}%
\bibitem [{\citenamefont {Pastawski}\ \emph {et~al.}(2015)\citenamefont
  {Pastawski}, \citenamefont {Yoshida}, \citenamefont {Harlow},\ and\
  \citenamefont {Preskill}}]{Pastawski15b}%
  \BibitemOpen
  \bibfield  {author} {\bibinfo {author} {\bibfnamefont {F.}~\bibnamefont
  {Pastawski}}, \bibinfo {author} {\bibfnamefont {B.}~\bibnamefont {Yoshida}},
  \bibinfo {author} {\bibfnamefont {D.}~\bibnamefont {Harlow}}, \ and\ \bibinfo
  {author} {\bibfnamefont {J.}~\bibnamefont {Preskill}},\ }\href {\doibase
  10.1007/JHEP06(2015)149} {\bibfield  {journal} {\bibinfo  {journal} {JHEP}\
  }\textbf {\bibinfo {volume} {06}},\ \bibinfo {pages} {149} (\bibinfo {year}
  {2015})}\BibitemShut {NoStop}%
\bibitem [{\citenamefont {Bouwmeester}\ \emph {et~al.}(1997)\citenamefont
  {Bouwmeester}, \citenamefont {Pan}, \citenamefont {Mattle}, \citenamefont
  {Eibl}, \citenamefont {Weinfurter},\ and\ \citenamefont
  {Zeilinger}}]{bouwmeester1997experimental}%
  \BibitemOpen
  \bibfield  {author} {\bibinfo {author} {\bibfnamefont {D.}~\bibnamefont
  {Bouwmeester}}, \bibinfo {author} {\bibfnamefont {J.-W.}\ \bibnamefont
  {Pan}}, \bibinfo {author} {\bibfnamefont {K.}~\bibnamefont {Mattle}},
  \bibinfo {author} {\bibfnamefont {M.}~\bibnamefont {Eibl}}, \bibinfo {author}
  {\bibfnamefont {H.}~\bibnamefont {Weinfurter}}, \ and\ \bibinfo {author}
  {\bibfnamefont {A.}~\bibnamefont {Zeilinger}},\ }\href@noop {} {\bibfield
  {journal} {\bibinfo  {journal} {Nature}\ }\textbf {\bibinfo {volume} {390}},\
  \bibinfo {pages} {575} (\bibinfo {year} {1997})}\BibitemShut {NoStop}%
\bibitem [{\citenamefont {Furusawa}\ \emph {et~al.}(1998)\citenamefont
  {Furusawa}, \citenamefont {S{\o}rensen}, \citenamefont {Braunstein},
  \citenamefont {Fuchs}, \citenamefont {Kimble},\ and\ \citenamefont
  {Polzik}}]{furusawa1998unconditional}%
  \BibitemOpen
  \bibfield  {author} {\bibinfo {author} {\bibfnamefont {A.}~\bibnamefont
  {Furusawa}}, \bibinfo {author} {\bibfnamefont {J.~L.}\ \bibnamefont
  {S{\o}rensen}}, \bibinfo {author} {\bibfnamefont {S.~L.}\ \bibnamefont
  {Braunstein}}, \bibinfo {author} {\bibfnamefont {C.~A.}\ \bibnamefont
  {Fuchs}}, \bibinfo {author} {\bibfnamefont {H.~J.}\ \bibnamefont {Kimble}}, \
  and\ \bibinfo {author} {\bibfnamefont {E.~S.}\ \bibnamefont {Polzik}},\
  }\href@noop {} {\bibfield  {journal} {\bibinfo  {journal} {Science}\ }\textbf
  {\bibinfo {volume} {282}},\ \bibinfo {pages} {706} (\bibinfo {year}
  {1998})}\BibitemShut {NoStop}%
\bibitem [{\citenamefont {Preskill}(2018)}]{preskill2018quantum}%
  \BibitemOpen
  \bibfield  {author} {\bibinfo {author} {\bibfnamefont {J.}~\bibnamefont
  {Preskill}},\ }\href@noop {} {\bibfield  {journal} {\bibinfo  {journal}
  {arXiv preprint arXiv:1801.00862}\ } (\bibinfo {year} {2018})}\BibitemShut
  {NoStop}%
\bibitem [{\citenamefont {Boixo}\ \emph {et~al.}(2016)\citenamefont {Boixo},
  \citenamefont {Isakov}, \citenamefont {Smelyanskiy}, \citenamefont {Babbush},
  \citenamefont {Ding}, \citenamefont {Jiang}, \citenamefont {Martinis},\ and\
  \citenamefont {Neven}}]{boixo2016characterizing}%
  \BibitemOpen
  \bibfield  {author} {\bibinfo {author} {\bibfnamefont {S.}~\bibnamefont
  {Boixo}}, \bibinfo {author} {\bibfnamefont {S.~V.}\ \bibnamefont {Isakov}},
  \bibinfo {author} {\bibfnamefont {V.~N.}\ \bibnamefont {Smelyanskiy}},
  \bibinfo {author} {\bibfnamefont {R.}~\bibnamefont {Babbush}}, \bibinfo
  {author} {\bibfnamefont {N.}~\bibnamefont {Ding}}, \bibinfo {author}
  {\bibfnamefont {Z.}~\bibnamefont {Jiang}}, \bibinfo {author} {\bibfnamefont
  {J.~M.}\ \bibnamefont {Martinis}}, \ and\ \bibinfo {author} {\bibfnamefont
  {H.}~\bibnamefont {Neven}},\ }\href@noop {} {\bibfield  {journal} {\bibinfo
  {journal} {arXiv preprint arXiv:1608.00263}\ } (\bibinfo {year}
  {2016})}\BibitemShut {NoStop}%
\bibitem [{\citenamefont {Renes}\ \emph {et~al.}(2004)\citenamefont {Renes},
  \citenamefont {Blume-Kohout}, \citenamefont {Scott},\ and\ \citenamefont
  {Caves}}]{renes2004symmetric}%
  \BibitemOpen
  \bibfield  {author} {\bibinfo {author} {\bibfnamefont {J.~M.}\ \bibnamefont
  {Renes}}, \bibinfo {author} {\bibfnamefont {R.}~\bibnamefont {Blume-Kohout}},
  \bibinfo {author} {\bibfnamefont {A.~J.}\ \bibnamefont {Scott}}, \ and\
  \bibinfo {author} {\bibfnamefont {C.~M.}\ \bibnamefont {Caves}},\ }\href@noop
  {} {\bibfield  {journal} {\bibinfo  {journal} {Journal of Mathematical
  Physics}\ }\textbf {\bibinfo {volume} {45}},\ \bibinfo {pages} {2171}
  (\bibinfo {year} {2004})}\BibitemShut {NoStop}%
\bibitem [{\citenamefont {Dankert}\ \emph
  {et~al.}(2009{\natexlab{a}})\citenamefont {Dankert}, \citenamefont {Cleve},
  \citenamefont {Emerson},\ and\ \citenamefont {Livine}}]{dankert2009exact}%
  \BibitemOpen
  \bibfield  {author} {\bibinfo {author} {\bibfnamefont {C.}~\bibnamefont
  {Dankert}}, \bibinfo {author} {\bibfnamefont {R.}~\bibnamefont {Cleve}},
  \bibinfo {author} {\bibfnamefont {J.}~\bibnamefont {Emerson}}, \ and\
  \bibinfo {author} {\bibfnamefont {E.}~\bibnamefont {Livine}},\ }\href@noop {}
  {\bibfield  {journal} {\bibinfo  {journal} {Physical Review A}\ }\textbf
  {\bibinfo {volume} {80}},\ \bibinfo {pages} {012304} (\bibinfo {year}
  {2009}{\natexlab{a}})}\BibitemShut {NoStop}%
\bibitem [{\citenamefont {Erd{\H o}s}\ and\ \citenamefont
  {Schr{\"o}der}(2014)}]{Erdos:2014aa}%
  \BibitemOpen
  \bibfield  {author} {\bibinfo {author} {\bibfnamefont {L.}~\bibnamefont
  {Erd{\H o}s}}\ and\ \bibinfo {author} {\bibfnamefont {D.}~\bibnamefont
  {Schr{\"o}der}},\ }\href {\doibase 10.1007/s11040-014-9164-3} {\bibfield
  {journal} {\bibinfo  {journal} {D. Math Phys Anal Geom}\ }\textbf {\bibinfo
  {volume} {17}},\ \bibinfo {pages} {441} (\bibinfo {year} {2014})}\BibitemShut
  {NoStop}%
\bibitem [{\citenamefont {Dankert}\ \emph
  {et~al.}(2009{\natexlab{b}})\citenamefont {Dankert}, \citenamefont {Cleve},
  \citenamefont {Emerson},\ and\ \citenamefont {Livine}}]{Dankert09}%
  \BibitemOpen
  \bibfield  {author} {\bibinfo {author} {\bibfnamefont {C.}~\bibnamefont
  {Dankert}}, \bibinfo {author} {\bibfnamefont {R.}~\bibnamefont {Cleve}},
  \bibinfo {author} {\bibfnamefont {J.}~\bibnamefont {Emerson}}, \ and\
  \bibinfo {author} {\bibfnamefont {E.}~\bibnamefont {Livine}},\ }\href@noop {}
  {\bibfield  {journal} {\bibinfo  {journal} {Phys. Rev. A}\ }\textbf {\bibinfo
  {volume} {80}},\ \bibinfo {pages} {012304} (\bibinfo {year}
  {2009}{\natexlab{b}})}\BibitemShut {NoStop}%
\end{thebibliography}%


\begin{thebibliography}{4}
\expandafter\ifx\csname natexlab\endcsname\relax\def\natexlab#1{#1}\fi
\expandafter\ifx\csname bibnamefont\endcsname\relax
  \def\bibnamefont#1{#1}\fi
\expandafter\ifx\csname bibfnamefont\endcsname\relax
  \def\bibfnamefont#1{#1}\fi
\expandafter\ifx\csname citenamefont\endcsname\relax
  \def\citenamefont#1{#1}\fi
\expandafter\ifx\csname url\endcsname\relax
  \def\url#1{\texttt{#1}}\fi
\expandafter\ifx\csname urlprefix\endcsname\relax\def\urlprefix{URL }\fi
\providecommand{\bibinfo}[2]{#2}
\providecommand{\eprint}[2][]{\url{#2}}

\bibitem[{\citenamefont{Kitaev}(2016)}]{Kitaev-Haar}
\bibinfo{author}{\bibfnamefont{A.}~\bibnamefont{Kitaev}}
  (\bibinfo{year}{2016}), \bibinfo{note}{course taught in spring 2016 at
  Caltech.}

\bibitem[{\citenamefont{Yoshida and Kitaev}(2017)}]{Yoshida:2017aa}
\bibinfo{author}{\bibfnamefont{B.}~\bibnamefont{Yoshida}} \bibnamefont{and}
  \bibinfo{author}{\bibfnamefont{A.}~\bibnamefont{Kitaev}}
  (\bibinfo{year}{2017}), \eprint{arXiv:1710.03363},
  \urlprefix\url{https://arxiv.org/abs/1710.03363}.

\bibitem[{\citenamefont{Beigi}(2013)}]{beigi2013sandwiched}
\bibinfo{author}{\bibfnamefont{S.}~\bibnamefont{Beigi}},
  \bibinfo{journal}{Journal of Mathematical Physics}
  \textbf{\bibinfo{volume}{54}}, \bibinfo{pages}{122202}
  (\bibinfo{year}{2013}).

\bibitem[{\citenamefont{Grover}(1996)}]{Grover:1996}
\bibinfo{author}{\bibfnamefont{L.~K.} \bibnamefont{Grover}}, in
  \emph{\bibinfo{booktitle}{Proceedings of the Twenty-eighth Annual ACM
  Symposium on Theory of Computing}} (\bibinfo{publisher}{ACM},
  \bibinfo{address}{New York, NY, USA}, \bibinfo{year}{1996}), STOC '96, pp.
  \bibinfo{pages}{212--219}, ISBN \bibinfo{isbn}{0-89791-785-5}.

\end{thebibliography}
\end{document}


\title{Supplemental Material for Disentangling Scrambling and Decoherence via Quantum Teleportation}
\author{Beni Yoshida and Norman Y.~Yao}
\maketitle

\section{Clifford scramblers versus Haar random unitaries}

The aforementioned $2$-qutrit  and $3$-qubit unitary operators are  Clifford operators which transform Pauli operators into Pauli operators. This is a rather restrictive class of unitary operators. For instance, a Haar random unitary $U$ will transform a Pauli operator $P$ into $UPU^{\dagger}=\sum_{Q \in \text{Pauli}}C_{Q}Q$ where $Q$ are Pauli operators, and $C_{Q}$ are  coefficients that are almost uniformly distributed over non-identity Pauli operators. Because of the special property of Clifford operators which preserve the Pauli group, the decoding task can be performed even without post-selection. To see this, let us explicitly consider the case with the $3$-qubit unitary operator. Bob performs measurements in the Bell basis:
\begin{align}
\frac{1}{\sqrt{2}}(|00\rangle + |11\rangle) \quad \frac{1}{\sqrt{2}}(|10\rangle + |01\rangle) \quad \frac{1}{\sqrt{2}}(|10\rangle - |01\rangle)  \quad \frac{1}{\sqrt{2}}(|00\rangle - |11\rangle) 
\end{align}
which can be also written as
\begin{align}
(I\otimes I)|\text{EPR}\rangle \quad (X\otimes I)|\text{EPR}\rangle \quad (Y\otimes I)|\text{EPR}\rangle \quad (Z\otimes I)|\text{EPR}\rangle.
\end{align}
The original probabilistic decoding protocol succeeded only when Bon measures $|\text{EPR}\rangle$ on $DD'$. Suppose that Bob had instead measured $(X\otimes I)|\text{EPR}\rangle$ instead of $|\text{EPR}\rangle$. The outcome can be represented graphically as follows
\begin{align}
\figbox{1.0}{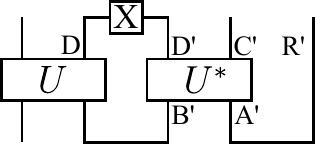}.
\end{align}
where a Pauli $X$ operator is inserted on a horizontal line connecting $DD'$. Since $U$ is a scrambling operator, there exists some operator $V$ supported on $C'R'$ which satisfies:
\begin{align}
\figbox{1.0}{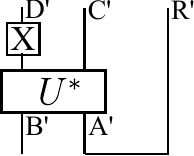}\approx \figbox{1.0}{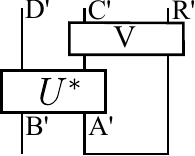}.
\end{align}
This statement follows from the fact that $I(A,BD)=I(D,AC)$ is nearly maximal. In general, the operator $V$ cannot be written as a tensor product of two operators acting on $C'$ and $R'$. In the case of $U$ being a Clifford operator, $V$ can be written as 
\begin{align}
V = P\otimes Q
\end{align}
where $P$ and $Q$ are some Pauli operators acting on $C'$ and $R'$ respectively. This implies that, by applying $Q$ on $R'$, one can reconstruct the original quantum state even if $(X\otimes I)|\text{EPR}\rangle$ was measured. It should be emphasized that this phenomena crucially relies on the fact that the time-evolution operator was a Clifford operator, and does not occur for generic scrambling unitary operators, such as a Haar random unitary. 

Another subtle difference between Haar random unitary operators and Clifford operators is the value of averaged OTOCs. Let us consider the case where $d_{A}\leq d_{D}$. If $U$ is drawn uniformly at random, the late-time asymptotic value is given by
\begin{align}
\langle \overline{\OTOC} \rangle \approx \frac{1}{d_{A}^2} + \frac{1}{d_{D}^2} - \frac{1}{d_{A}^2d_{D}^2}.
\end{align}
Namely, the values of $\langle \overline{\OTOC} \rangle$ for different random unitary operators will not differ much as the variance is suppressed by the system size. On the other hand, if $U$ is drawn from Clifford operators, we have the same ensemble average
\begin{align}
\int_{U \in \text{Clifford}} dU\ \langle \overline{\OTOC} \rangle \approx \frac{1}{d_{A}^2} + \frac{1}{d_{D}^2} - \frac{1}{d_{A}^2d_{D}^2}.
\end{align}
However, the statistical variance of $\langle \overline{\OTOC} \rangle$  is not suppressed by the system size. The reason why this variance remains unsuppressed is simple: the values of OTOCs taken with respect to Pauli operators, are either $\pm1$, and become small only after taking an average over $O_{A}$ and $O_{D}$. Thus, according to our fine-grained definition of scrambling, random Clifford operators are not scrambling as four-point OTOCs do not decompose as in Eqn.~(1). On the other hand, such random Clifford operators  do satisfy  our coarse-grained definition of average scrambling.

Finally, Clifford operations are typically assumed to be (relatively) easy to implement  while non-Clifford gates are significantly more challenging. Let us consider a scenario where we would like to check if a given unitary operator $U$ is a Clifford operator or not. One approach, which is motivated by OTOCs, is to measure the commutator
\begin{align}
\langle P(t)Q(0)P(t)Q(0) \rangle = \frac{1}{d}\text{Tr} ( UPU^{\dagger}Q UPU^{\dagger}Q)
\end{align}
for randomly chosen Pauli operators $P,Q$. If $U$ is a Clifford operator, then the above quantity should be either $+1$ or $-1$. The merit of this method is that one can tell if $U$ is Clifford or not after only a few trials with reasonable confidence via relatively simple operations.


\section{Towards a finite temperature generalization}

We have treated the cases where the quantum state is maximally mixed in OTOCs; $\langle O_{X}O_{Y}(t)O_{Z}O_{W}(t)\rangle = \Tr( O_{X}O_{Y}(t)O_{Z}O_{W}(t) \frac{1}{d} \mathbb{I})$. The relation between the mutual information $I(A,BD)$ and OTOCs can be generalized to cases where the input and output ensembles factorize; $\rho_{AB}=\rho_{A}\otimes \rho_{B}$ and $\rho_{CD}=\rho_{C}\otimes \rho_{D}$. This generalization was originally discussed in~\cite{Kitaev-Haar,Yoshida:2017aa}. Note that previously, at infinite temperature, $\rho_{A}=\frac{1}{d_{A}}\mathbb{I}_{A}$, $\rho_{B}=\frac{1}{d_{B}}\mathbb{I}_{B}$, $\rho_{C}=\frac{1}{d_{C}}\mathbb{I}_{D}$, $\rho_{D}=\frac{1}{d_{D}}\mathbb{I}_{D}$. 
%
%
 In order to consider the state representation of a unitary operator $U$ with the initial quantum state $\rho_{AB}$, we simply replace each dot with $\rho_{R}^{1/2}$ on the corresponding Hilbert space $R$ in Eqn.~(21):
\begin{align}
\figbox{1.0}{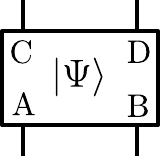} =  \figbox{1.0}{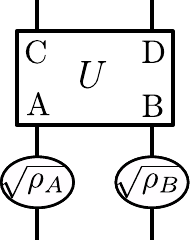}=  \figbox{1.0}{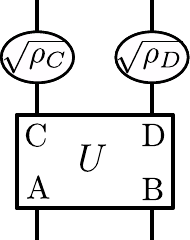}.
\end{align}
The goal is to lower bound the mutual information $I(A,BD)$ of the above pure state $|\Psi\rangle$ from OTOCs. To do so, we think of preparing the thermofield double states for $\rho_{A}$ on $RA$ and $A'R'$, and the thermofield double state for $\rho_{B}$ on $BB'$. Note that the thermofield double state can be created by replacing a dot in the EPR pair with $\rho_{R}^{1/2}$ on the corresponding Hilbert space $R$. We then performs a projective measurement onto the thermofield double state for $\rho_{D}$ on $DD'$. Defining the density matrix $\rho=|\Psi\rangle\langle \Psi|$, the amplitude for this projective measurement is given by 
\begin{align}
P = \Tr \Big[   \rho_{BD} (\rho_{B}\otimes \rho_{D}^{-1})^{-1/2}  \rho_{BD}  (\rho_{B}\otimes \rho_{D}^{-1})^{-1/2} \Big].
\end{align}
Using the sandwiched R\'{e}nyi-$\alpha$ divergence, defined as $D_{\alpha}(f||g)=\frac{1}{\alpha - 1} \log \left(  \frac{1}{\text{Tr} (f)} \text{Tr} \Big[\big( g^{\frac{1-\alpha}{2\alpha}}f g^{\frac{1-\alpha}{2\alpha}}\big)^{\alpha} \Big] \right)$, the amplitude can be written as 
\begin{align}
\log_{2} P = D_{2}(\rho_{BD} | \rho_{B}\otimes \rho_{D}^{-1} )  \label{eq:P-divergence}
\end{align}
Using the monotonicity of R\'{e}nyi-$\alpha$ divergence~\cite{beigi2013sandwiched}, it can be bounded as follows:
\begin{align}
\log_{2}(P) \geq D_{1}(\rho_{BD} | \rho_{B}\otimes \rho_{D}^{-1} ) = -S_{BD} + S_{B} - S_{D}.
\end{align}
The righthand side is equal to $- I(A,BD)$ due to unitarity of $U$, so we have
\begin{align}
I(A,BD) \geq - \log_{2} P.
\end{align}
The rest is to relate the amplitude $P$ to OTOCs. One may consider the following two types of OTOCs:
\begin{equation}
\begin{split}
\langle O_{X}O_{Y}(t)O_{Z}O_{W}(t)\rangle_{1} =\Tr( O_{X}O_{Y}(t)O_{Z}O_{W}(t) \rho_{AB})\\
\langle O_{X}O_{Y}(t)O_{Z}O_{W}(t)\rangle_{2} =\Tr( O_{X}O_{Y}(t) \sqrt{\rho_{AB}} O_{Z}O_{W}(t) \sqrt{\rho_{AB}})
\end{split}
\end{equation}
where subscripts correspond to ``one-sided'' or ``two-sided'' geometries of a black hole. It is not difficult to see that the amplitude $P$ can be expressed as a certain weighted average of the two-sided OTOCs $\langle O_{X}O_{Y}(t)O_{Z}O_{W}(t)\rangle_{2}$. One may consider a similar weighted average of the one-sided OTOCs $\langle O_{X}O_{Y}(t)O_{Z}O_{W}(t)\rangle_{1}$ which upper bounds the average of the two-sided OTOCs $\langle O_{X}O_{Y}(t)O_{Z}O_{W}(t)\rangle_{2}$ via the Cauchy-Schwarz inequality. Therefore, smallness of certain averaged OTOCs, either one-sided or two-sided, is sufficient to lower bound $I(A,BD)$.

\section{Deterministic decoder with Grover search}

In the main part of the paper, we have discussed a probabilistic decoding protocol which works with probability $\approx \frac{1}{d_{A}^2}$ where $d_{A}$ is the size of the Hilbert space for Alice's message. In this appendix, we briefly describe a deterministic decoding protocol which incorporates a procedure similar to the Grover search algorithm~\cite{Grover:1996}. The circuit complexity of the deterministic decoding protocol is proportional to $d_{A}$ for large $d_{A}$, and the whole process is related to higher-point OTOCs. For a qubit input ($d_{A}=2$), the protocol requires only one iteration, and the whole process is related to six-point and eight-point OTOCs. 

\begin{figure}
\centering\includegraphics[width=0.7\linewidth]{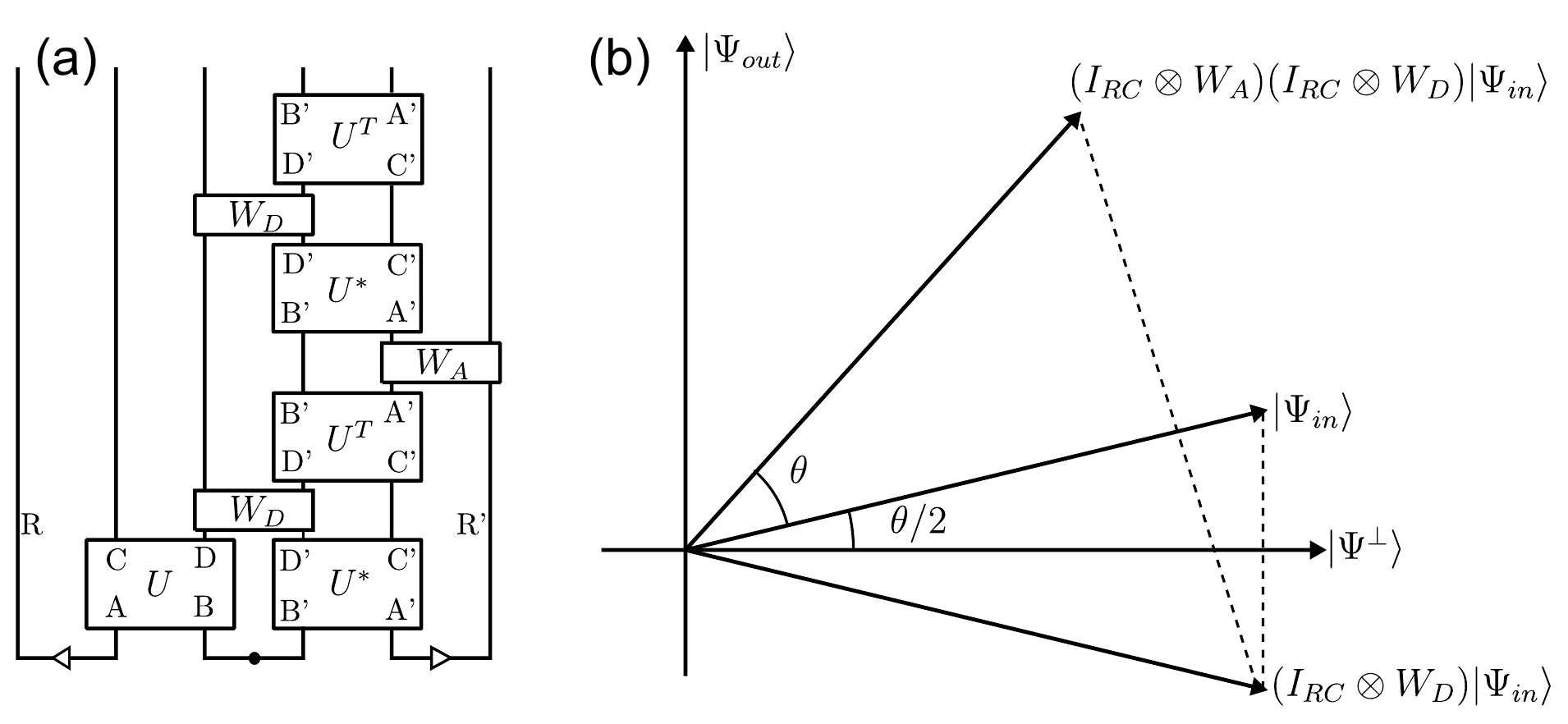}
\caption{Schematic depiction of the the deterministic decoder (a) and the Grover rotation (b).}
\label{Figure-Grover}
\end{figure}

The initial state of the protocol is 
\begin{align}
|\Psi_{\text{in}}\rangle 
= \: \figbox{1.0}{fig-in-state}\:, \label{eq-in-state}
\end{align}
where Bob has already applied $U^{*}$ to his share of qubits. Define the following unitary operators
\begin{align}
W_{D} = 1 - 2 (I_{RC}\otimes P_{D}) \qquad W_{A} = 2 (I_{RC}\otimes \widetilde{P}_{A})-1
\end{align}
where $P_{D}$ is a projector onto EPR pairs on $DD'$, $P_{A}$ is a projector onto EPR pairs on $A'R'$, and $\widetilde{P}_{A}=(I_{D}\otimes U^{*}\otimes I_{R'})P_{A}(I_{D}\otimes U^{T}\otimes I_{R'})$, or graphically:
\begin{align}
P_{A} = \figbox{1.0}{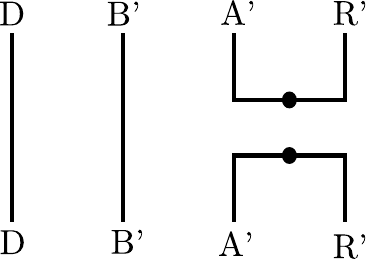}, \qquad\quad \widetilde{P_{A}} = \figbox{1.0}{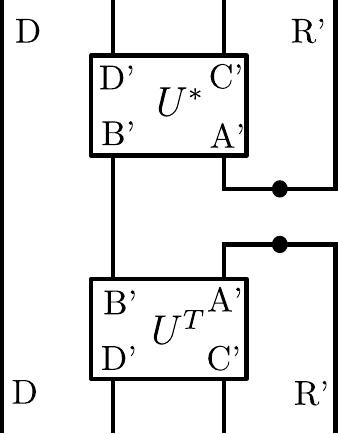}\,\:.
\end{align}
Bob's decoding strategy is to implement a unitary operator $V\equiv V_{A}V_{D}$ multiple ($\approx \frac{\pi d_{A} }{4}$) times to obtain a good approximation of $|\Psi_{\text{out}}\rangle$. The protocol is summarized in Fig.~\ref{Figure-Grover}(a). 

To illustrate how the protocol works, we will use some equations that hold in the ideal case with $I^{(2)}(A,BD)=\log_{2}d_{A}^2$: 
\begin{equation}
\begin{split}
I_{RC}\otimes P_{D}|\Psi_{\text{in}}\rangle = \frac{1}{d_{A}}|\Psi_{\text{out}}\rangle \qquad I_{RC}\otimes  P_{D}|\Psi_{\text{out}}\rangle= |\Psi_{\text{out}}\rangle \\
I_{RC}\otimes \widetilde{P}_{A}|\Psi_{\text{in}}\rangle = |\Psi_{\text{in}}\rangle \qquad I_{RC}\otimes \widetilde{P}_{A} |\Psi_{\text{out}}\rangle = \frac{1}{d_{A}}|\Psi_{\text{in}}\rangle. \label{eq:PAPD}
\end{split}
\end{equation}
The rest is the standard analysis of the Grover search algorithm. Consider a two-dimensional plane spanned by $|\Psi_{\text{in}}\rangle$ and $|\Psi_{\text{out}}\rangle$ with real coefficients. Notice that applications of $W_{A},W_{D}$ keep wavefunctions on the two-dimensional plane. Let $|\Psi_{\perp}\rangle $ be a wavefunction which lies on this plane and is orthogonal to $|\Psi_{\text{out}}\rangle$; $\langle\Psi_{\perp}|\Psi_{\text{out}}\rangle=0$. Such a wavefunction can be constructed by observing $|\Psi_{\perp}\rangle\propto(1-P_{D})|\Psi_{\text{out}}\rangle$. Notice that $V_{D}$ is a reflection across $|\Psi_{\perp}\rangle$, so this induces a rotation by angle $\theta$ with $\sin \frac{\theta}{2} = \frac{1}{d_{A}}$ when applied to $|\Psi_{\text{in}}\rangle$. Similarly, $W_{A}$ is a reflection across $|\Psi_{\text{in}}\rangle$ (see Fig.~\ref{Figure-Grover}(b)). Therefore, by applying $W=W_{A}W_{D}$, one can rotate $|\Psi_{\text{in}}\rangle$ on the two-dimensional plane by angle $\theta$. After $m$ steps, we have 
\begin{align}
|\Psi(m)\rangle = \sin\Big(\big(m+\frac{1}{2}\big)\theta \Big)|
\Psi_{\text{out}}\rangle + 
\cos\Big(\big(m+\frac{1}{2}\big)\theta \Big)|
\Psi_{\perp}\rangle.
\end{align}
So, the probability of obtaining $|\Psi_{\text{out}}\rangle$ is $\sin^{2} \Big( \big( m+ \frac{1}{2} \big)\theta \Big)$. 

When $I^{(2)}(A,BD)\approx 2\log_{2}d_{A}$ and $d_{A}=2$ (the input is a qubit), an almost perfect decoding is possible by applying $U^{*}$ and $U^{T}$. This is because $\theta=\pi/3$. Namely, the following iteration implements the deterministic decoding:
\begin{align}
\figbox{1.0}{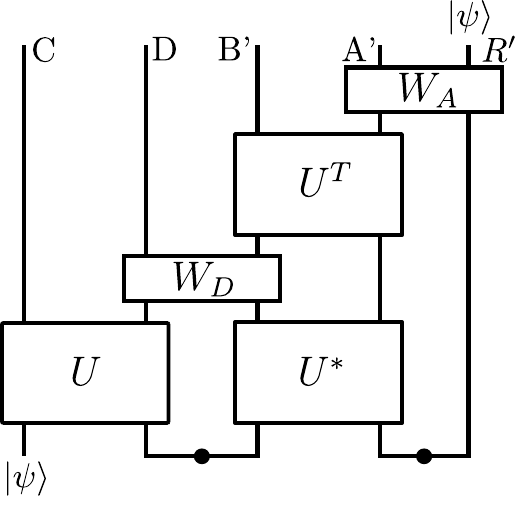}
\end{align}
If one further applies $U^{*}$ on the right hand side and $W_{D}$ on $DD'$, we will have an EPR pair on $DD'$: 
\begin{align}
\figbox{1.0}{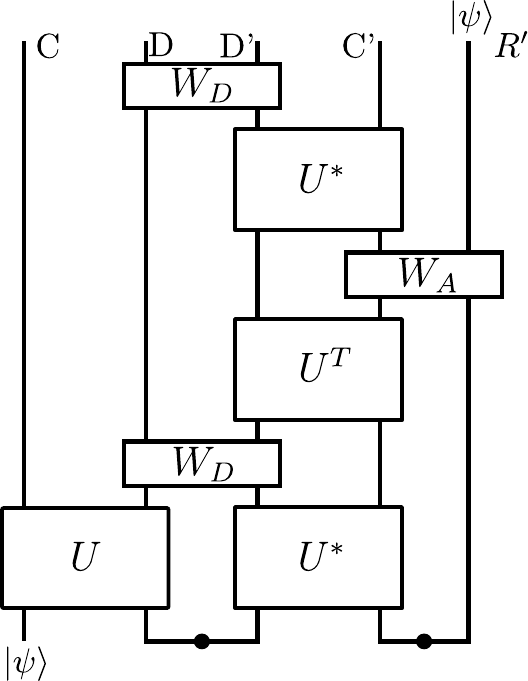}
\end{align}
One may postselect the experiment by using an EPR pair on $DD'$.

\mciteSetMidEndSepPunct{}{\ifmciteBstWouldAddEndPunct.\else\fi}{\relax}
%
\bibliography{reference2018Jan}{}